\documentclass[12pt,a4]{article}

\usepackage{color,tikz}
\usepackage[unicode,bookmarks,bookmarksopen,bookmarksopenlevel=2,colorlinks,linkcolor=blue,citecolor=green]{hyperref}

\usepackage{amsmath,eucal,amssymb,amsthm,amsfonts}
\usepackage{mathrsfs,graphicx,texdraw}

\newtheorem{remark}{Remark}

\def\im{{\mbox{Im}}}

\def\ad{\mathrm{ad\,}}
\def\openone{\leavevmode\hbox{\small1\kern-3.3pt\normalsize1}}

\def\bbbc{{\Bbb C}}
\def\bbbr{{\Bbb R}}

\def\a{{\boldsymbol a}}
\def\b{{\boldsymbol b}}

\def\c{{\boldsymbol c}}
\def\d{{\boldsymbol d}}

\def\T{{\boldsymbol T}}
\def\S{{\boldsymbol S}}

\def\v{{\boldsymbol v}}
\def\w{{\boldsymbol w}}

\def\r{{\boldsymbol r}}
\def\s{{\boldsymbol s}}
\def\p{{\boldsymbol p}}
\def\q{{\boldsymbol q}}

\begin{document}

\begin{center}
{\LARGE \bf On integrable wave interactions and  Lax pairs\\[4pt] on symmetric spaces}

\bigskip

{\bf Vladimir S. Gerdjikov$^{a,b,c,}$\footnote{E-mail: {\tt gerjikov@inrne.bas.bg}},
Georgi  G. Grahovski$^{a,d,}$\footnote{E-mail: {\tt grah@essex.ac.uk}} and
Rossen  I. Ivanov$^{e,}$\footnote{E-mail: {\tt rivanov@dit.ie}}}

\end{center}

\medskip

\noindent
{\it $^{a}$ Institute for Nuclear Research and Nuclear Energy,
Bulgarian Academy of Sciences, 72 Tsarigradsko chaussee, 1784 Sofia, BULGARIA }\\
{\it $^b$Institute of Mathematics and Informatics, Bulgarian Academy of Sciences,
8 Georgi Bonchev Str.,  1113 Sofia, Bulgaria}\\
{\it $^c$ Institute for Advanced Physical Studies, New Bulgarian University, 21 Montevideo str. Sofia 1618, Bulgaria}\\
{\it $^{d}$ Department of Mathematical Sciences, University of Essex, Wivenhoe Park, Colchester CO4 3SQ, UK}\\
{\it $^{e}$ School of Mathematical Sciences, Dublin Institute of Technology, Kevin Street, Dublin 8, IRELAND }

\begin{abstract}
\noindent
Multi-component generalizations of derivative nonlinear Schr\"odinger (DNLS) type of equations having quadratic bundle Lax pairs
related to $\mathbb{Z}_2$-graded Lie algebras and
 {\bf A.III} symmetric spaces are studied.  The  Jost solutions and the minimal set of scattering data for the case of local and nonlocal
reductions are constructed. The latter lead to multi-component integrable equations with ${\cal CPT}$-symmetry. Furthermore,
the fundamental analytic solutions (FAS) are constructed and the spectral properties of the associated Lax operators
are briefly discussed. The Riemann-Hilbert problem (RHP) for the multi-component generalizations of DNLS equation of
Kaup-Newell (KN) and Gerdjikov-Ivanov (GI) types is derived.  A modification of the dressing method is presented allowing the
explicit derivation of  the soliton solutions for the multi-component GI equation
with both local and nonlocal reductions. It is shown that for specific choices of the reduction these solutions
can have  regular behavior for all finite $x$ and $t$. The fundamental properties
  of the multi-component GI equations are briefly discussed at the end.
\end{abstract}

\tableofcontents

\section{Introduction}

\noindent
The generalized Zakharov-Shabat systems:
\begin{equation}\label{eq:zsh}\begin{split}
L\psi(x,t,\lambda)\equiv  i \frac{\partial \psi}{ \partial x } + (Q(x,t) - \lambda J)\psi(x,t,\lambda)=0,
\end{split}\end{equation}
and the associated nonlinear evolutionary equations (NLEE) have attracted  considerable attention of the mathematical and
physical communities over the last four decades and have been rather well classified and analyzed \cite{ZMNP,67,FaTa, GVYa*08,DokLeb,GK*81}.
This class of NLEE contains such physically important equations as the nonlinear Schr\"oodinger equation (NLS), the sine-Gordon, the
$2$-dimensional Toda chain and modified Korteweg–de-Vriez (mKdV) equations. All these models are integrable by the Inverse Scattering
Method (ISM). Here $Q(x,t)$ is a matrix-valued function belonging in general to a simple Lie algebra ${\frak g}$ of rank $n$, $J$ is an (constant)
element of its Cartan subalgebra ${\frak h} \subset {\frak g}$ and $\lambda \in {\Bbb C}$ is a spectral parameter.

Another classical example of completely integrable NLEE is the derivative nonlinear Schr\"o\-din\-ger (DNLS) equation  (also known as Kaup-Newell equation) \cite{KN,76}:
\begin{equation}\label{eq:DNLS}
  {\rm i}q_t+q_{xx}+\epsilon {\rm i}(|q|^2q)_x=0.
\end{equation}
This equation has many physical applications, especially in plasma physics and nonlinear optics \cite{Ruder1,Ruder2,Mio,Mjo}: it  describes small-amplitude nonlinear Alfv\'en waves in
a low-$\beta$ plasma, propagating strictly parallel or at a small angle to the ambient
magnetic field; it models also large-amplitude
magnetohydrodynamic (MHD) waves in a high-$\beta$ plasma propagating at an
arbitrary angle to the ambient magnetic field.
Here the subscripts $x$ and $t$ stay for partial derivatives with respect to the variables $x$ and $t$ respectively and $\epsilon =\pm 1$. It is related to a
Lax pair quadratic in $\lambda$ \cite{KN}:
\begin{eqnarray}
 L(\lambda) &=& {\rm i}\partial_x +\lambda Q(x,t)-\lambda^2\sigma_3; \label{eq:LaxL-DNLS}\\
  M(\lambda) &=& {\rm i}\partial_t + \sum_{k=1}^3M_k(x,t)\lambda^k -2\lambda^4\sigma_3\label{eq:LaxM-DNLS}
\end{eqnarray}
where
\begin{eqnarray}
  Q(x,t) &=& \left(\begin{array}{cc}
                     0 & q(x,t) \\
                     \epsilon q^*(x,t) & 0
                   \end{array}
  \right), \qquad \sigma_3=\left(\begin{array}{cc}
                     1 & 0 \\
                     0 & -1
                   \end{array}
  \right), \qquad M_3(x,t)=2Q(x,t) \label{eq:Pot-DNLS}\\
  M_2 (x,t) &=& \epsilon |q^2(x,t)|\sigma_3, \qquad M_1(x,t)={{\rm i}\over 2}[\sigma_3Q_x(x,t)]+\epsilon  |q^2(x,t)|Q(x,t).
\end{eqnarray}
DNLS is closely related (via  gauge transformations) to three other integrable NLEEs: the one studied by Chen-Lee-Liu \cite{Lee}
\begin{equation}\label{eq:Chen}
  {\rm i}q_t+q_{xx}+{\rm i}|q|^2q_x=0,
\end{equation}
and the equation studied by Gerdjikov-Ivanov (GI) \cite{20,21,Fan,Yilmaz}
\begin{equation}\label{eq:GI}
  {\rm i}q_t+q_{xx}+\epsilon {\rm i}q^2q_x^* + {1\over 2}|q|^4q(x,t)=0,
\end{equation}
and the 2-dimensional Thirring model \cite{KuzMikh}. Due to their similarity with DNLS, (\ref{eq:Chen}) and (\ref{eq:GI}) are sometimes
termed DNLS II and DNLS III, respectively. All the three versions of the DNLS equation along with the 2-dimensional Thirring
model are integrable by the ISM and are related to spectral problems for generic quadratic bundle Lax operators (related to the algebra $sl(2, {\Bbb C})$)
\begin{equation}\label{eq:QuadBundle}
L(\lambda)={\rm i}\partial_x + U_0(x,t) + \lambda U_1(x,t) -\lambda^2\sigma_3,
\end{equation}
where $U_0(x,t)$ is a  generic ($2\times 2$) traceless matrix, $U_1(x,t)$ is an off-diagonal matrix and $\sigma_3$
being the Pauli matrix. The $N$-th AKNS flow of the corresponding integrable hierarchy \cite{AKNS,GVYa*08,ZMNP} is given by a second Lax operator $M(\lambda)$ of a polynomial form:
\begin{equation}\label{eq:QuadBundle2}
M(\lambda)= i \partial_t + \sum_{k=0}^NV_k(x,t)\lambda^k.
\end{equation}
A very fruitful trend in the theory of integrable systems is the study of multi-component generalizations of integrable scalar NLEEs. The applications of the differential geometric and Lie algebraic
methods to soliton type equations led to the discovery of a close
relationship between the multi-component (matrix) integrable equations (of nonlinear Schr\"odinger type)
and the symmetric and homogeneous spaces~\cite{ForKu*83}.  Later on, this approach was extended to other types
of multi-component integrable models, like the derivative NLS,
Korteweg--de~Vries and modified Korteweg--de~Vries, $N$-wave,
Davey--Stewartson, Kadomtsev--Petviashvili equations
\cite{AthFor,For}. For example, the equation \cite{For} (see also \cite{TV0,TV2})
\begin{equation}\label{eq:Fordy}
{\rm i}{\bf q}_t + {\bf q}_{xx} + {2m{\rm i}\over m + n}({\bf q}{\bf q}^\dag {\bf q})_x = 0,
\end{equation}
is related to symmetric space ${\bf A.III}\simeq SU(m + n)/S(U(m) \times U(n))$ in the Cartan classification \cite{Helg}.
Here ${\bf q}(x,t)$ is a smooth $n \times m$ matrix-valued function and ``$\dag$''   stands for Hermitian conjugation. Similarly, Tsuchida and Wadati \cite{Tsuchida} proved
the complete integrability of a class of matrix generalizations of the Chen-Lee-Liu equation \cite{Lee}.

Another important trend in the development of ISM was the introduction of the reduction group
by A. V. Mikhailov \cite{2}, and further developed in \cite{vgrn,GGK*01,GGK*01b,GGK05a,Grah}. This allows
one to prove that some of the well known integrable models  and also a number
of new interesting NLEE   are integrable by the ISM and possess special symmetry
properties. As a result  the potential $Q(x, t)$ has a very special form imposed by the reduction. The reduction group concept is important
also because of the fact, that when one considers  Lax operators on Lie algebras, the number of independent fields grows rather quickly with
the rank of the algebra: the corresponding NLEEs are solvable for any rank, but their possible
applications to physics do not seem realistic. However, one still may extract new integrable and
physically useful NLEE by imposing reductions on $L(\lambda) $, i.e. algebraic restrictions on the potential
of L, which diminish the number of independent functions in them and the number of equations. Of course, such restrictions must be compatible with the dynamics of the NLEE.

Along with the well-known local reductions, recently nonlocal integrable reductions gained a fast-growing interest: in \cite{AblMus} was proposed  a nonlocal integrable equation of nonlinear Schr\"odinger type with
${\cal PT}$-symmetry, due to the invariance of
the so-called self-induced potential $V(x,t)=\psi(x,t)\psi^*(-x,-t)$ under the combined action of parity and time
reversal symmetry (\ref{eq:PT}). In the same paper, the 1-soliton solution for this model is derived and it was shown
that it develops  singularities in finite time. Soon after this, nonlocal ${\cal PT}$-symmetric generalizations are found for
 the Ablowitz-Ladik model in \cite{AblMus1}. All these models are integrable by the Inverse Scattering Method (ISM) \cite{AblMus2}. To be fair, we should note that in
 the context of the theory of integrable systems such nonlocal reductions appear for the first time in studies of integrable boundary conditions \cite{Chered,Sklyan,Khabib}.

${\cal PT}$-symmetric systems gained a special interest in the last decade mainly due to their applications in Nonlinear Optics \cite{UFN,Bender1,Bender2, Ali1,Ali2,Nature}.

The initial interest in such systems was motivated by quantum mechanics \cite{Bender1, Ali1}. In \cite{Bender1}
it was shown that quantum systems with a non-Hermitian Hamiltonian admit states with real eigenvalues, i.e. the Hermiticity
of the Hamiltonian is not a necessary condition to have  real spectrum. Using such Hamiltonians one can build  up new quantum mechanics
\cite{Bender1,Bender2, Ali1,Ali2}.  Starting point is the fact that in the case of a non-Hermitian Hamiltonian with real spectrum,
the modulus of the wave function for the eigenstates is time-independent even in the case of complex potentials.

Historically the first pseudo-Hermitian Hamiltonian with real spectrum is the ${\cal PT}$-symmetric one in \cite{Bender1}. Pseudo-Hermiticity here means that
the Hamiltonian ${\cal H}$ commutes with the operators of spatial reflection ${\cal P}$ and time reversal ${\cal T}$: ${\cal P}{\cal T}{\cal H}={\cal H}{\cal P}{\cal T}$.
The action of these operators is defined as follows: ${\cal P}: x\to -x$ and ${\cal T}: t\to -t$.

Supposing that the wave function is a scalar, this leads to the following action of the operator of spatial reflection on the space of states:
\begin{equation}\label{eq:P}
{\cal P}\psi(x,t)=\psi(-x,t).
\end{equation}
Applying similar arguments to the (anti-linear and anti-unitary) time reversal operator ${\cal T}$ shows that it should act on the space of states as follows:
\begin{equation}\label{eq:T}
{\cal T}\psi(x,t)=\psi^*(x,-t).
\end{equation}
Therefore, the Hamiltonian and the wave function are ${\cal PT}$-symmetric, if
\begin{equation}\label{eq:PT}
{\cal H}(x,t)={\cal H}^*(-x,-t), \qquad \psi(x,t)=\psi^*(-x,-t).
\end{equation}
In addition if one imposes charge conjugation symmetry (particle-antiparticle symmetry) ${\cal C}$ \cite{Peskin}, this will lead to additional complex conjugation of the wave function and the Hamiltonian:
\begin{equation}\label{eq:C}
{\cal CH}^*(x,t)={\cal H}(x,t), \qquad {\cal C}\psi^*(x,t)=\psi(x,t).
\end{equation}
One can show that the ${\cal C}$-symmetry can be realized by a unitary linear operator \cite{Peskin}. The Hamiltonian and the wave function are ${\cal CPT}$-symmetric, if
\begin{equation}\label{eq:CPT}
{\cal H}(x,t)={\cal H}(-x,-t), \qquad \psi(x,t)=\psi(-x,-t).
\end{equation}
Integrable systems with ${\cal PT}$-symmetry were studied extensively over the last two decades \cite{Fring1,Fring2,Konotop,Barash,Barash1,GGK*XX}.

The main purpose of the present paper is to study generic quadratic in $\lambda$ Lax operators
\begin{equation}\label{eq:QB}
L(\lambda) = {\rm i}\partial_x +\lambda Q(x,t)-\lambda^2J,
\end{equation}
related to $\mathbb{Z}_2$-graded Lie algebras and symmetric spaces of ${\bf A.III}$-type and the corresponding multi-component generalizations of DNLS III
or Gerdjikov-Ivanov (GI) equation. This includes: the spectral properties of the associated Lax operator, the direct scattering transform, the
effect of reductions on the scattering data, the dressing method, soliton solutions and related Riemann-Hilbert problem.
We also aim to study the effect of local and nonlocal reductions on the multi-component generalizations of the GI equation.

The paper is organized as follows: In Section 2 we outline the construction of Lax pairs related to symmetric spaces of {\bf A.III}-type
and the related NLEE, the direct scattering problem for the corresponding Lax operator: this includes the construction of the Jost solutions and
the minimal set of scattering data. In Section 3 we describe the local and nonlocal reductions and their action on the Lax operators. The construction of
the fundamental analytic solutions (FAS) is outlined in Section 4. In Section 5 we construct the Riemann-Hilbert problem for the Kaup-Newell (KN) and
GI equations on {\bf A.III}-symmetric spaces and derive explicit parametrization of the associated Lax operators. The effect of local and nonlocal reductions on the scattering data is described briefly in Section 6. The modification of the dressing method and
1-solitons for the multi-component GI equation with local and nonlocal reductions are presented in Section 7. The integrals of
motion  are briefly discussed in Section 8. We finish up with conclusions and a brief outlook in Section 9.

\section{Preliminaries}

\subsection{NLEE related to graded Lie algebras and symmetric spaces}

We will start with NLEE related to $\mathbb{Z}_2$-graded Lie algebras, see \cite{Helg}:
\begin{equation}\label{eq:g0}\begin{split}
\mathfrak{g} \simeq  \mathfrak{g}^{(0)} \oplus  \mathfrak{g}^{(1)} , \qquad  \mathfrak{g}^{(0)} \equiv
\{  X\in  \mathfrak{g}, \; [J, X]=0\}, \qquad \mathfrak{g}^{(1)} \equiv \{  Y\in  \mathfrak{g}, \; J Y+Y J=0\},
\end{split}\end{equation}
where $J = \left(\begin{array}{cc} \openone_p & 0 \\ 0 & -\openone_q \end{array}\right)$.
More specifically in the typical representation of the algebra $\mathfrak{g}\simeq sl(n)$ with $n=p+q$ the subalgebra
$\mathfrak{g}^{(0)}$ consists of block-diagonal matrices with two nontrivial blocks $p\times p$ and $q\times q$; the linear subspace
$\mathfrak{g}^{(1)}$ consists of block-off-diagonal matrices with two nontrivial blocks $p\times q$ and $q\times p$.

 We can also say that $\mathfrak{g}^{(1)}$ is the co-adjoint orbit passing through $J$ which will play role of phase space for
 our NLEE.

 \begin{remark}\label{rem:1}
In fact,  the grading (\ref{eq:g0}) can be related to the tangent hyperplane to $SL(p+q)/(SL(p)\otimes SL(q))$. If we introduce in
addition a complex structure on $\mathfrak{g}$ the grading (\ref{eq:g0}) can be related to the Hermitian symmetric space
${\bf A.III}\simeq SU(p+q)/(S(U(p)\otimes U(q))$.
 \end{remark}

As we mentioned above we will be considering Lax operators that are given by:
\begin{equation}\label{eq:L2}\begin{split}
 L \psi &\equiv i \frac{\partial \psi }{ \partial x } + (U_2(x,t) + \lambda Q(x,t) -\lambda^2 J) \psi(x,t,\lambda) =0,
 \qquad Q(x,t) = \left(\begin{array}{cc} 0 & \q \\ \p & 0   \end{array}\right), \\
 M \psi &\equiv i \frac{\partial \psi }{ \partial t } + (V_4(x,t) + \lambda V_3(x,t) + \lambda^2 V_2(x,t)
 +  \lambda^3 Q(x,t)-\lambda^4 J) \psi(x,t,\lambda) =0.
\end{split}\end{equation}
where $Q(x,t)$,  $   V_3(x,t) \in \mathfrak{g}^{(1)}$ and $U_2(x,t)$,  $ V_2(x,t)$ and  $ V_4(x,t) \in \mathfrak{g}^{(0)}$.
Such Lax pairs give rise to multicomponent derivative NLS type equations.
The Lax pair (\ref{eq:L2}), as we shall see below, allows one to solve the system of NLEE:
\begin{equation}\label{eq:NLEE2*}\begin{split}
i \frac{\partial \q}{ \partial t} + \frac{1}{2} \frac{\partial^2 \q }{ \partial x^2 } - \frac{i}{2} \q \frac{\partial \p}{ \partial x } \q + \frac{1}{4} \q \p \q \p \q &=0, \\
-i \frac{\partial \p}{ \partial t} + \frac{1}{2} \frac{\partial^2 \p }{ \partial x^2 } + \frac{i}{2} \p \frac{\partial \q}{ \partial x } \p + \frac{1}{4} \p \q \p \q \p &=0.
\end{split}\end{equation}
Along with it we will consider the Lax pair:
\begin{equation}\label{eq:L2t}\begin{split}
\tilde{L} \tilde{ \psi} &\equiv i \frac{\partial \tilde{ \psi} }{ \partial x } +  (\lambda \tilde{ Q}(x,t) -\lambda^2 J)\tilde{ \psi}(x,t,\lambda) =0,
 \qquad \tilde{ Q}(x,t) = \left(\begin{array}{cc} 0 & \tilde{ \q} \\ \tilde{ \p} & 0   \end{array}\right), \\
 \tilde{ M} \tilde{ \psi} &\equiv i \frac{\partial \tilde{\psi} }{ \partial t } + ( \lambda \tilde{V}_3(x,t) + \lambda^2 \tilde{V}_2(x,t)
 +  \lambda^3 \tilde{Q}(x,t)-\lambda^4 J)\tilde{ \psi}(x,t,\lambda) =0.
\end{split}\end{equation}
which is gauge equivalent to (\ref{eq:L2}) and which allows one to solve the system:
\begin{equation}\label{eq:KNm*}\begin{split}
i \frac{\partial \tilde{\q}}{ \partial t } &+ \frac{\partial^2 \tilde{\q} }{ \partial x^2 } + i \frac{\partial \tilde{\q}\tilde{\p}\tilde{\q}}{ \partial x }=0, \\
-i \frac{\partial \tilde{\p}}{ \partial t } &+ \frac{\partial^2 \tilde{\p} }{ \partial x^2 } - i \frac{\partial \tilde{\p}\tilde{\q}\tilde{\p}}{ \partial x }=0.
\end{split}\end{equation}
The second system also complies with the $\mathbb{Z}_2$-grading introduced above:
$\tilde{ Q}(x,t)$,  $   \tilde{ V}_3(x,t) \in \mathfrak{g}^{(1)}$ and $\tilde{ U}_2(x,t)$ and  $\tilde{ V}_2(x,t) \in \mathfrak{g}^{(0)}$.

Note that the use of $\mathbb{Z}_2$-grading is in fact additional reduction as compared to the block-matrix AKNS model
\cite{AKNS,GVYa*08}. Indeed, the block-matrix AKNS method requires  that only $Q(x,t)\in \mathfrak{g}^{(1)}$.

\subsection{The scattering problem for $L $}\label{ssec:2.1}

Here we briefly outline the scattering problem for the system (\ref{eq:L2}) for the class of potentials $Q(x,t) $ satisfying
the following conditions:

{\bf  C1:}  $Q(x,t) $ is smooth enough and falls off to zero fast enough for $x\to\pm\infty  $ for all $t $.

{\bf  C2:}  $Q(x,t) $ is such that $L $ has at most  finite number of simple discrete eigenvalues.

In this subsection $t $  plays the role of an additional parameter; for the sake of brevity the $t $-dependence is not
always shown. The condition C2 cannot be formulated as a set of explicit conditions on $Q(x,t) $; its precise meaning will become
clear below.

The main tool here is the Jost solutions defined by their asymptotics at $x\to\pm\infty  $:
\begin{equation}\label{eq:5.1}
\lim_{x\to\infty } \psi (x,\lambda )e^{i \lambda^2 Jx} =\openone , \qquad \lim_{x\to -\infty } \phi (x,\lambda )e^{i \lambda^2 Jx}
=\openone ,
\end{equation}
Along with the Jost solutions, we introduce
\begin{equation}\label{eq:5.1a}
X_+ (x,\lambda ) =\psi (x,\lambda )e^{i \lambda^2 Jx}, \qquad X_- (x,\lambda ) =\phi (x,\lambda )e^{i \lambda^2 Jx};
\end{equation}
which satisfy the following linear integral equations
\begin{eqnarray}\label{eq:5.2}
X_\pm (x,\lambda ) &= \openone + i \int_{\pm \infty }^{x} d y e^{-i \lambda^2 J(x-y)} Q(y) X_\pm (y,\lambda ) e^{i \lambda^2 J(x-y)}.
\end{eqnarray}
These are Volterra type equations which, as is well known always have solutions provided one can ensure the convergence of the
integrals on the right hand side. For real $\lambda  $  the exponential factors in (\ref{eq:5.2})  are just
oscillating and the convergence is ensured by condition C1.

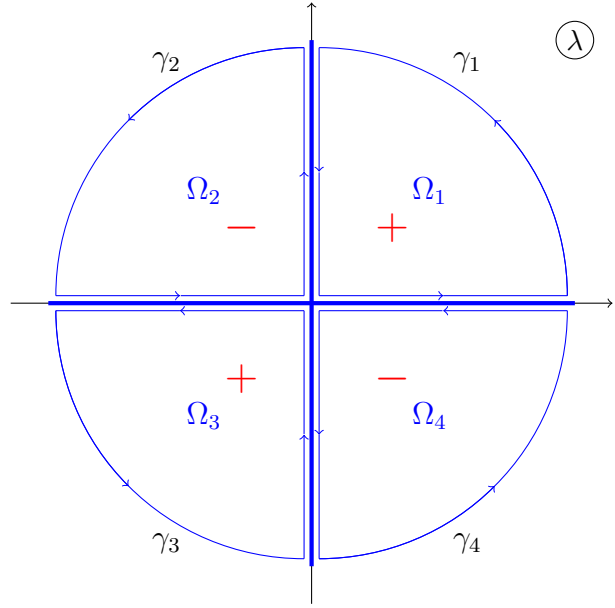
\begin{figure}
\begin{center}
\begin{tikzpicture}

\draw [->,black](0,-4) -- (0,4);
\draw [->,black](-4,0) -- (4,0);

\draw [-,ultra thick, blue](-3.5,0) -- (3.5,0);
\draw [-,ultra thick, blue](0,-3.5) -- (0,3.5);

\draw [black] (3.5,3.5) node  {{\small $\lambda$}};

\draw [black] (3.5,3.5) circle (0.25);

\draw [black] (2,3.2) node  {{  $\gamma_1$}};
\draw [black] (-2,3.2) node  {{  $\gamma_2$}};
\draw [black] (-2,-3.2) node  {{  $\gamma_3$}};
\draw [black] (2,-3.2) node  {{  $\gamma_4$}};

\draw [red] (1.0,1.0) node  {{  \Large $+$}};
\draw [blue] (1.5,1.5) node  {{  $\Omega_1$}};
\draw [red] (-1.0,1.0) node  {{ \Large $-$}};
\draw [blue] (-1.5,1.5) node  {{  $\Omega_2$}};
\draw [blue] (-1.5,-1.5) node  {{  $\Omega_3$}};
\draw [red] (-1.0,-1.0) node  {{ \Large $+$}};
\draw [blue] (1.5,-1.5) node  {{  $\Omega_4$}};
\draw [red] (1.0,-1.0) node  {{ \Large $-$}};

\draw [-, blue](3.4,0.1) arc (0:90:3.3);
\draw [->, blue](3.4,0.1) arc (0:45:3.3);
\draw [->,blue](0.1,0.1) -- (1.75,0.1);
\draw [-,blue](1.75,0.1) -- (3.4,0.1);
\draw [->,blue](0.1,3.4) -- (0.1,1.75);
\draw [-,blue](0.1,1.74) -- (0.1,0.1);

\draw [-, blue,rotate=90](3.4,0.1) arc (0:90:3.3);
\draw [->, blue,rotate=90](3.4,0.1) arc (0:45:3.3);
\draw [->,blue,rotate=90](0.1,0.1) -- (1.75,0.1);
\draw [-,blue,rotate=90](1.75,0.1) -- (3.4,0.1);
\draw [->,blue,rotate=90](0.1,3.4) -- (0.1,1.75);
\draw [-,blue,rotate=90](0.1,1.74) -- (0.1,0.1);

\draw [-, blue,rotate=180](3.4,0.1) arc (0:90:3.3);
\draw [->, blue,rotate=180](3.4,0.1) arc (0:45:3.3);
\draw [->,blue,rotate=180](0.1,0.1) -- (1.75,0.1);
\draw [-,blue,rotate=180](1.75,0.1) -- (3.4,0.1);
\draw [->,blue,rotate=180](0.1,3.4) -- (0.1,1.75);
\draw [-,blue,rotate=180](0.1,1.74) -- (0.1,0.1);

\draw [-, blue,rotate=270](3.4,0.1) arc (0:90:3.3);
\draw [->, blue,rotate=270](3.4,0.1) arc (0:45:3.3);
\draw [->,blue,rotate=270](0.1,0.1) -- (1.75,0.1);
\draw [-,blue,rotate=270](1.75,0.1) -- (3.4,0.1);
\draw [->,blue,rotate=270](0.1,3.4) -- (0.1,1.75);
\draw [-,blue,rotate=270](0.1,1.74) -- (0.1,0.1);

\end{tikzpicture}
\end{center}
\caption{The continuous spectrum of a $L(\lambda)$. }
\label{fig:1}
\end{figure}

The Jost solutions as whole cannot be extended for $\im \lambda^2 \neq 0 $.  Skipping the details we write down the Jost
solutions $\psi (x,\lambda ) $ and $\phi (x,\lambda ) $ in the following block-matrix form:
\begin{equation}\label{eq:5.3}
\psi (x,\lambda ) = \left(|\psi^- (x,\lambda )\rangle , |\psi^+ (x,\lambda )\rangle \right), \qquad  \phi (x,\lambda ) =
\left(|\phi^+ (x,\lambda ) \rangle , |\phi^- (x,\lambda )\rangle \right),
\end{equation}
where the superscripts $+ $ and (resp. $- $) show that the corresponding block-matrix allows analytic extension
for $\lambda \in \Omega_1\cup \Omega_3 $ (resp. $\lambda \in \Omega_2\cup \Omega_4 $),
see Figure \ref{fig:1}. Here by $\Omega_1, \dots, \Omega_4 $ we have denoted the quadrants of the
complex $\lambda$-plane.

Solving the direct scattering problem means given the potential $Q(x) $ to find the scattering matrix $T(\lambda ) $. By
definition $T(\lambda ) $ relates the two Jost solutions:
\begin{equation}\label{eq:6.1}
\phi (x,\lambda ) =\psi (x,\lambda )T(\lambda ), \qquad T(\lambda )= \left(\begin{array}{cc} \a^+(\lambda ) & -\b^-(\lambda )\\
\b^+(\lambda ) & \a^-(\lambda ) \end{array}\right)
\end{equation}
and has compatible block-matrix structure.  In what follows we will need also the inverse of the scattering matrix
(`hat'  means:inverse matrix from now on):
\begin{equation}\label{eq:6.2i}
\psi (x,\lambda ) =\phi (x,\lambda )\hat{T}(\lambda ), \qquad \hat{T}(\lambda )\equiv \left(\begin{array}{cc} \c^-(\lambda ) &
\d^-(\lambda ) \\ -\d^+(\lambda ) & \c^+(\lambda ) \end{array}\right),
\end{equation}
where
\begin{equation}\label{eq:6.2}\begin{split}
\c^-(\lambda ) &= \hat{\a}^+(\lambda ) (\openone +\rho ^-\rho ^+)^{-1} = (\openone +\tau^+\tau^-)^{-1} \hat{\a}^+(\lambda ), \\
\d^-(\lambda ) &= \hat{\a}^+(\lambda )\rho ^- (\lambda ) (\openone  +\rho ^+\rho ^-)^{-1} = (\openone +\tau^+ \tau^-)^{-1} \tau^+ (\lambda ) \hat{\a}^-(\lambda ), \\
 \c^+(\lambda ) &= \hat{\a}^-(\lambda ) (\openone +\rho^+\rho ^-)^{-1}= (\openone +\tau^- \tau^+)^{-1}\hat{\a}^-(\lambda ), \\
\d^+(\lambda ) &= \hat{\a}^-(\lambda )\rho ^+(\lambda ) (\openone  +\rho ^-\rho ^+)^{-1}= (\openone +\tau^-\tau^+)^{-1} \tau^- (\lambda )\hat{\a}^+(\lambda ).
\end{split}\end{equation}
The diagonal blocks of both $T(\lambda ) $ and $\hat{T}(\lambda )$  allow analytic continuation off the real axis, namely
$\a^+(\lambda ) $, $\c^+(\lambda ) $ are analytic functions of $\lambda  $ for $\lambda \in \Omega_1\cup \Omega_3 $, while  $\a^-(\lambda )
$, $\c^-(\lambda ) $  are analytic functions of $\lambda$ for $\lambda \in \Omega_2\cup \Omega_4 $.

By $\rho ^\pm(\lambda ) $ and $\tau^\pm(\lambda ) $ above we have denoted the multicomponent generalizations of the reflection
coefficients (for the scalar case, see \cite{AKNS,KN}):
\begin{equation}\label{eq:rho-tau}
\rho ^\pm(\lambda ) =\b^\pm\hat{\a}^\pm (\lambda ) =\hat{\c}^\pm \d^\pm(\lambda ), \qquad \tau^\pm(\lambda ) =\hat{\a}^\pm
\b^\mp(\lambda ) =\d^\mp\hat{\c}^\pm (\lambda ) ,
\end{equation}
We will need also the asymptotics  for $\lambda \to\infty  $:
\begin{equation}\label{eq:6.2b}\begin{split}
 \lim_{\lambda \to -\infty } \phi (x,\lambda ) e^{i \lambda Jx} = \lim_{\lambda \to\infty } \psi (x,\lambda ) e^{i \lambda Jx}
=\openone , \qquad \lim_{\lambda \to\infty } T(\lambda )  =\openone ,
\end{split}\end{equation}
i.e. $\lim_{\lambda \to\infty } \a^\pm (\lambda )  = \lim_{\lambda \to\infty } \c^\pm(\lambda )  = \openone$.

The inverse to the Jost solutions $\hat{\psi }(x,\lambda) $ and $\hat{\phi }(x,\lambda) $ are solutions to:
\begin{equation}\label{eq:L-inv}
i {d \hat{\psi }\over d x } - \hat{\psi }(x,\lambda) (U_2(x,t) + \lambda Q(x,t)-\lambda^2 J) =0,
\end{equation}
satisfying the conditions:
\begin{equation}\label{eq:J-inv}
\lim_{x\to\infty } e^{-i \lambda Jx}\hat{\psi}(x,\lambda)=\openone , \qquad \lim_{x\to -\infty } e^{-i \lambda
Jx}\hat{\phi }(x,\lambda)=\openone .
\end{equation}
Now it is the collections of rows of $\hat{\psi }(x,\lambda) $ and $\hat{\phi }(x,\lambda) $ that possess analytic properties in
$\lambda  $:
\begin{equation}\label{eq:psi-inv}
\hat{\psi }(x,\lambda) = \left( \begin{array}{c} \langle \hat{\psi }^+(x,\lambda) | \\ \langle \hat{\psi }^-(x,\lambda) |
\end{array} \right), \qquad
\hat{\phi }(x,\lambda) = \left( \begin{array}{c} \langle \hat{\phi }^-(x,\lambda) | \\ \langle \hat{\phi }^+(x,\lambda) | \end{array} \right),
\end{equation}
Just like the Jost solutions, their inverse (\ref{eq:psi-inv}) are solutions to linear equations (\ref{eq:L-inv}) with regular
boundary conditions (\ref{eq:J-inv}); therefore they can have no singularities in their regions of analyticity. The same holds true
also for the scattering matrix $T(\lambda )=\hat{\psi }(x,\lambda) \phi (x,\lambda) $ and its inverse $\hat{T}(\lambda )=\hat{\phi
}(x,\lambda) \psi (x,\lambda) $, i.e.
\begin{equation}\label{eq:a-pm}
\a^+(\lambda ) = \langle \hat{\psi }^+(x,\lambda) |\phi^+(x,\lambda) \rangle , \qquad \a^-(\lambda ) = \langle \hat{\psi
}^-(x,\lambda) |\phi^- (x,\lambda) \rangle ,
\end{equation}
as well as
\begin{equation}\label{eq:c-pm}
\c^+(\lambda ) = \langle \hat{\phi }^+(x,\lambda) |\psi^+(x,\lambda) \rangle , \qquad \c^-(\lambda ) = \langle \hat{\phi
}^-(x,\lambda) |\psi^- (x,\lambda) \rangle ,
\end{equation}
are analytic for $\lambda \in \bbbc_\pm $ and have no singularities in their regions of analyticity.  However they may
become degenerate (i.e., their determinants may vanish) for some values $\lambda _j^\pm \in \bbbc_\pm $ of $\lambda  $. Below we
analyze the structure of these degeneracies.

\section{Local and Non-local Reductions}\label{ssec:2.3}

\subsection{Local Reductions}

An important and systematic tool to construct new integrable NLEE is the so-called  reduction group \cite{2}.
It will be instructive to start with the local reductions:
\begin{equation}\label{eq:Z2-Mi}\begin{aligned}
& \mbox{1)}& \qquad A_1U^{\dagger}(x,t,\kappa _1\lambda ^*)A_1^{-1} &= U(x,t,\lambda ),
&\quad A_1V^{\dagger}(x,t,\kappa _1\lambda ^*)A_1^{-1} &= V(x,t,\lambda ), \\
 & \mbox{2)} & \qquad A_2U^{T}(x,t,\kappa _2\lambda )A_2^{-1} &= -U(x,t,\lambda ), &\quad
A_2V^{T}(x,t,\kappa _2\lambda )A_2^{-1} &= -V(x,t,\lambda ), \\
& \mbox{3)}& \qquad A_3U^{*}(x,t,\kappa _1\lambda ^*)A_3^{-1} &= -U(x,t,\lambda ), &\quad
A_3V^{*}(x,t,\kappa _1\lambda^* )A_3^{-1} &= -V(x,t,\lambda ), \\
 &\mbox{4)}& \qquad A_4U(x,t,\kappa _2\lambda )A_4^{-1} &= U(x,t,\lambda ), &\quad
A_4V(x,t,\kappa _2\lambda )A_4^{-1} &= V(x,t,\lambda ).
\end{aligned}\end{equation}
The consequences of these reductions and the constraints they impose on the FAS and the Gauss
factors of the scattering matrix are well known, see \cite{2,ZMNP,67,GGK*01,vgrn}.

Let us detail the consequences of the reductions 1) and 3) in (\ref{eq:Z2-Mi}) on the NLEE (\ref{eq:NLEE2*}).
It is easy to see that they restrict $U_0(x,t) $ and $Q(x,t)$ by:
\begin{equation}\label{eq:rest1}\begin{aligned}
& \mbox{1)}  & \quad A_1JA_1^{-1} &= J, &\quad \kappa_1 A_1 Q^\dag A_1^{-1}& = Q(x,t), &\quad A_1 U_2^\dag (x,t) A_1^{-1} &= U_2(x,t), \\
& \mbox{3)}  & \quad A_3JA_3^{-1} &= -J, &\quad \kappa_3 A_3 Q^* A_3^{-1} &=- Q(x,t), &\quad A_3 U_2^* (x,t) A_3^{-1} &= -U_2(x,t),
\end{aligned}\end{equation}
where $\kappa_1^2 = \kappa_3^2=1$ and $A_1^2 = A_3^2=\openone$. From $A_1JA_1^{-1} = J$ (resp. $A_3JA_3^{-1} = -J$)
we find that $A_1$ is block-diagonal (resp. $A_3$ is block-off-diagonal) matrix. If we introduce
\begin{equation}\label{eq:A13}\begin{split}
A_1 = \left(\begin{array}{cc} a_1 & 0 \\ 0 & a_2  \end{array}\right) , \qquad  A_3 = \left(\begin{array}{cc} 0 & b_1 \\ b_2 & 0 \end{array}\right) ,
\end{split}\end{equation}
we obtain
\begin{equation}\label{eq:pq-dag}\begin{aligned}
& \mbox{1)}  & \quad  \kappa_1 a_1 \p^\dag \hat{a}_2 & = \q, &\quad  \kappa_1 a_2 \q^\dag \hat{a}_1 & = \p, & \quad A_1 U_2^\dag A_1^{-1} & =U_2 \\
& \mbox{3)}  & \quad  \kappa_3 b_1 \p^* \hat{b}_2 & = -\q, &\quad  \kappa_3 b_2 \q^* \hat{b}_1 & = -\p, & \quad A_3 U_2^* A_3^{-1} & =-U_2 .
\end{aligned}\end{equation}
As a result the eq. (\ref{eq:NLEE2*}) reduces to a multicomponent GI equation:
\begin{equation}\label{eq:mGI}\begin{split}
i \frac{\partial \q}{ \partial t} + \frac{1}{2} \frac{\partial^2 \q }{ \partial x^2 } - \frac{i \kappa_1}{2}  \q a_2 \frac{\partial \q^\dag}{ \partial x } \hat{a}_1
\q + \frac{1}{4} \q a_2 \q^\dag \hat{a}_1 \q a_2 \q^\dag \hat{a}_1 \q =0.
\end{split}\end{equation}
while the equation (\ref{eq:KNm*}) goes into a multicomponent KN equation:
\begin{equation}\label{eq:KN*}\begin{split}
i \frac{\partial \tilde{\q}}{ \partial t }  + \frac{\partial^2 \tilde{\q} }{ \partial x^2 } + i \kappa_1 \frac{\partial }{ \partial x } \left(
\tilde{\q} a_2 \tilde{\q}^\dag \hat{a}_1 \tilde{\q}\right) =0.
\end{split}\end{equation}


\subsection{Non-Local Reductions}

It is important to note, that for the derivative NLS equations {\em there are no }  reductions
compatible with either $\mathcal{P}$- or  $\mathcal{T}$-symmetry
separately. However the ${\Bbb Z}_2$ reductions
\begin{equation}\label{eq:Z2-nl}\begin{aligned}
& \mbox{1)}& \;\; C_1U^{\dagger}(-x, -t ,\kappa _1\lambda^* )C_1^{-1} &= -U(x,t,\lambda ),
&\;C_1V^{\dagger}(-x,- t ,\kappa _1\lambda^* )C_1^{-1} &= -V(x,t,\lambda ), \\
 & \mbox{2)} & \;\; C_2U^{T}(-x,- t ,\kappa _2\lambda )C_2^{-1} &= U(x,t,\lambda ), &\; C_2V^{T}(-x, -t , \kappa _2\lambda )C_2^{-1} &= V(x,t,\lambda ), \\
& \mbox{3)}& \;\; C_3U^{*}(-x, -t ,\kappa _1\lambda^*)C_3^{-1} &= U(x,t,\lambda ), &\;C_3V^{*}(-x, -t ,\kappa _1\lambda^*)C_3^{-1} &= V(x,t,\lambda ), \\
 &\mbox{4)}& \;\; C_4U(-x,- t ,\kappa _2\lambda )C_4^{-1}&= -U(x,t,\lambda ), &\;C_4V(-x, -t ,\kappa _2\lambda )C_4^{-1} &= -V(x,t,\lambda ),
\end{aligned}\end{equation}
are obviously ${\cal PT}$-symmetric \cite{TV}. Here $\kappa_i^2 =1$ and  $A_i$ and $C_i$, $i=1,\dots, 4$ are involutive automorphisms of the relevant Lie algebra.

Now the consequences of the reductions 1) and 3) in (\ref{eq:Z2-nl}) on the NLEE (\ref{eq:NLEE2*}).
It is easy to see that they restrict $U_0(x,t) $ and $Q(x,t)$ by:
\begin{equation}\label{eq:rest1'}\begin{aligned}
& \mbox{1)}  & \quad C_1JC_1^{-1} &= -J, &\quad \kappa_1 C_1 Q^\dag(-x,-t) C_1^{-1}& = -Q(x,t), &\quad C_1 U_2^\dag (-x,-t) C_1^{-1} &= -U_2(x,t), \\
& \mbox{3)}  & \quad C_3JC_3^{-1} &= J, &\quad \kappa_3 C_3 Q^*(-x,-t) C_3^{-1} &= Q(x,t), &\quad C_3 U_2^* (-x,-t) C_3^{-1} &=U_2(x,t),
\end{aligned}\end{equation}
where $\kappa_1^2 = \kappa_3^2=1$ and $C_1^2 = C_3^2=\openone$. From $C_1JC_1^{-1} = -J$ (resp. $C_3JC_3^{-1} = J$)
we find that $C_3$ is block-diagonal (resp. $C_1$ is block-off-diagonal) matrix. If we introduce
\begin{equation}\label{eq:C13}\begin{split}
C_1 = \left(\begin{array}{cc} 0 & c_1  \\ c_2 & 0  \end{array}\right) , \qquad  C_3 = \left(\begin{array}{cc} d_1 & 0 \\ 0 &  d_2  \end{array}\right) ,
\end{split}\end{equation}
we obtain
\begin{eqnarray}\label{eq:pq-dag'}
 \mbox{1)} &&  \; \kappa_1 c_1 \q^\dag (-x,-t)\hat{c}_2  = -\q(x,t) , \qquad
\kappa_1 c_2 \p^\dag (-x,-t)\hat{c}_1  = -\p(x,t), \nonumber\\
&&C_1 U_2^\dag(-x,-t) C_1^{-1}  =-U_2(x,t) \\
\mbox{3)}  && \; \kappa_3 d_1 \q^*(-x,-t) \hat{d}_2  = \q(x,t), \qquad
\kappa_3 d_2 \p^*(-x,-t) \hat{d}_1  = \p(x,t), \nonumber\\
&&C_3 U_2^*(-x,-t) C_3^{-1}  =U_2(x,t) .\nonumber
\end{eqnarray}
As a result the equations (\ref{eq:NLEE2*}) and (\ref{eq:KNm*}) retain their form, the only difference being that $\q$ and $\p$ are now restricted
by (\ref{eq:pq-dag'}).

On the Jost solutions we have
\begin{equation*}\label{eq:nlocR2}\begin{split}
\phi ^\dag (x,t,\lambda^*) = \psi^{-1}  (-x,t,-\lambda) , \qquad  \psi ^\dag (x,t,\lambda^*) = \phi^{-1}  (x,t,-\lambda) ,
\end{split}\end{equation*}
so for the scattering matrix we have
\begin{equation*}\label{eq:Tdag'}\begin{split}
 T ^\dag (t,-\lambda^*) = T  (t,\lambda) ,
\end{split}\end{equation*}
As a consequence for the Gauss factors we get:
\begin{equation*}\label{eq:Spm'}\begin{aligned}
 T^-{}^\dag (-\lambda^*) & = \hat{S}^+(\lambda), &\qquad  T^+{}^\dag (-\lambda^*) & = \hat{S}^-(\lambda), \qquad
 D^\pm{}^\dag (\lambda^*) & = \hat{D}^\pm(-\lambda).
\end{aligned}\end{equation*}
In analogy with the local reductions, the kernel of the resolvent has poles at the points $\lambda_2^\pm$ at which
$D^\pm (\lambda)$ has poles or zeroes. In particular, if $\lambda_2^+$ is an eigenvalue, then $-\lambda_2^+$ is also an
eigenvalue. For the reflection coefficients we obtain the constraints:
\begin{equation*}\label{eq:tau'}\begin{split}
\tau^+(-\lambda) = -\rho^{+,*} (\lambda), \qquad  \tau^-(-\lambda) = -\rho^{-,*} (\lambda),
\end{split}\end{equation*}

\begin{remark}\label{rem:2}
 In what follows for the sake of simplicity we specify $A_1=C_3=J$ and $A_3=C_1=\left(\begin{array}{cc} 0 & \openone \\
 \openone & 0   \end{array}\right)$. In the latter case we restrict ourselves to the special case when $\p$ and $\q$ are square
 matrices, i.e. our symmetric space is $SU(2q)/S(U(q)\otimes U(q))$.
\end{remark}

\section{The fundamental analytic solutions and the RHP}\label{ssec:FAS}

The next step is to construct the fundamental analytic solutions of (\ref{eq:L2}). In our case this is done simply by combining
the blocks of Jost solutions with the same analytic properties:
\begin{equation}\label{eq:6.3}\begin{split}
\chi ^+(x,\lambda ) &\equiv  \left(|\phi ^+\rangle , |\psi ^+\rangle\right)(x,\lambda) = \phi (x,\lambda ) \S^+(\lambda ) =
\psi (x,\lambda ) \T^-(\lambda ), \\
\chi ^-(x,\lambda ) &\equiv  \left(|\psi ^-\rangle , |\phi ^-\rangle \right)(x,\lambda ) = \phi (x,\lambda ) \S^-(\lambda ) =
\psi (x,\lambda ) \T^+(\lambda ),
\end{split}\end{equation}
where the block-triangular functions $\S^\pm(\lambda ) $ and
$\T^\pm(\lambda ) $ are given by:
\begin{equation}\label{eq:6.4}\begin{aligned}
\S^+(\lambda ) &= \left( \begin{array}{cc} \openone  & \d^-(\lambda ) \\ 0 & \c^+(\lambda ) \end{array}\right), &\qquad
\T^-(\lambda ) &=  \left( \begin{array}{cc} \a^+(\lambda ) & 0 \\ \b^+(\lambda ) & \openone  \end{array}\right),  \\
\S^-(\lambda ) &=  \left( \begin{array}{cc} \c^-(\lambda ) & 0 \\ -\d^+(\lambda ) & \openone  \end{array}\right), &\qquad
\T^+(\lambda ) &= \left( \begin{array}{cc} \openone  & -\b^-(\lambda )\\ 0 & \a^-(\lambda ) \end{array}\right),
\end{aligned}\end{equation}
These triangular factors can be viewed also as generalized Gauss decompositions (see \cite{Helg}) of $T(\lambda ) $ and its inverse:
\begin{equation}\label{eq:7.1}\begin{split}
T(\lambda ) &= \T^-(\lambda )\hat{\S}^+(\lambda ) =\T^+(\lambda )\hat{\S}^-(\lambda ), \qquad
\hat{T}(\lambda ) = \S^+(\lambda )\hat{\T}^-(\lambda ) =\S^-(\lambda )\hat{\T}^+(\lambda ).
\end{split}\end{equation}
The relations between $\c^\pm(\lambda ) $, $\d^\pm(\lambda ) $ and $\a^\pm(\lambda ) $, $\b^\pm(\lambda ) $ in eq. (\ref{eq:6.2})
ensure that equations (\ref{eq:7.1}) become identities. From eqs. (\ref{eq:6.3}), (\ref{eq:6.4}) we derive:
\begin{equation}\label{eq:9.4}\begin{aligned}
\chi ^+(x,\lambda ) &= \chi ^-(x,\lambda ) G_0(\lambda ), &\qquad G_0(\lambda ) &= \hat{D}^-(\lambda )(\openone +K^-(\lambda )),  \\
 \chi ^-(x,\lambda ) &= \chi ^+(x,\lambda ) \hat{G}_0(\lambda ),  &\qquad \hat{G}_0(\lambda ) &=\hat{D}^+(\lambda )(\openone -K^+(\lambda )),
\end{aligned}\end{equation}
valid for $\lambda \in \bbbr$, where
\begin{equation}\label{eq:K-pm}\begin{aligned}
D^-(\lambda ) &= \left( \begin{array}{cc} \c^-(\lambda ) &0 \\ 0 & \a^-(\lambda ) \end{array}\right), &\qquad
K^-(\lambda ) &= \left( \begin{array}{cc} 0& \d^-(\lambda ) \\ \b^+(\lambda ) & 0\end{array}\right), \\
D^+(\lambda ) &= \left( \begin{array}{cc} \a^+(\lambda ) &0 \\ 0 & \c^+(\lambda ) \end{array}\right), &\qquad
K^+(\lambda ) &= \left( \begin{array}{cc} 0& \b^-(\lambda ) \\ \d^+(\lambda ) & 0\end{array}\right),
\end{aligned}\end{equation}
Obviously the block-diagonal factors $D^+(\lambda ) $ and $D^-(\lambda )$ are matrix-valued analytic functions for $\lambda
\in \Omega_1\cup \Omega_3 $ and  $\lambda \in \Omega_2\cup \Omega_4 $ respectively. Another well known fact about
the FAS $\chi ^\pm (x,\lambda ) $ concerns their asymptotic behavior for $\lambda \to\pm\infty  $, namely:
\begin{equation}\label{eq:12.1}
\xi^\pm(x,\lambda ) = \chi ^\pm(x,\lambda ) e^{i \lambda^2 Jx}, \qquad
\lim_{\lambda \to\infty } \xi^\pm(x,\lambda ) =\openone .
\end{equation}
On the real and imaginary axis $\xi ^+(x, \lambda ) $ and $\xi ^-(x, \lambda ) $ are related by
\begin{equation}\label{eq:1.10}
\xi ^+(x,\lambda ) =\xi ^-(x,\lambda ) G(x,\lambda ), \qquad G(x,\lambda ) = e^{-i \lambda^2 Jx}G_0(\lambda) e^{i \lambda^2 Jx},
\qquad G_0(\lambda )= S^+(\lambda )\hat{S}^-(\lambda ).
\end{equation}
The function $G_0(\lambda ) $ can be considered as a minimal set of scattering data in the case of absence of discrete eigenvalues of
(\ref{eq:L2}) \cite{Sh,IP2}.

Thus eq. (\ref{eq:1.10}) combined with eq. (\ref{eq:12.1}) can be understood as a Riemann-Hilbert problem with
canonical normalization: given the sewing function $G_0(x,\lambda)$ construct $\xi^\pm(x,\lambda)$.

\section{Parametrization of Lax pairs}

Here we will outline a natural parametrization of $U(x,t,\lambda)$ and $V(x,t,\lambda)$ in terms of the
local coordinate  $Q_1(x,t)$ on the co-adjoint orbit $\mathfrak{g}^{(1)}$. Below we will choose it in the form:
\begin{equation}\label{eq:Q1}\begin{split}
 Q_1(x,t) = \frac{1}{2}\left(\begin{array}{cc} 0 & \q \\ -\p & 0   \end{array}\right),
\end{split}\end{equation}
where $\q$ and $\p$ are generic $p\times q$ and $q\times p$ matrices. Following \cite{DriSok,107s} we also
introduce the solution $\xi(x,t,\lambda)$ of a RHP with canonical normalization. Since $\xi(x,t,\lambda)$  must
be an element of the corresponding Lie group we define it by
\begin{equation}\label{eq:xi01}\begin{split}
\xi (x,t,\lambda) & = \exp (\mathcal{Q}(x,t,\lambda)), \qquad
\mathcal{Q}(x,t,\lambda)= \sum_{s=1}^{\infty} \lambda^{-s} Q_{s}(x,t),
\end{split}\end{equation}
where
$\mathcal{Q}(x,t,\lambda)$ is a formal series over the negative powers of $\lambda$ whose coefficients
$Q_s$ take values in $\mathfrak{g}^{(0)}$ if $s$ is even and in $\mathfrak{g}^{(1)}$ if $s$ is odd. Therefore
the first few of these coefficients take the form:
\begin{equation}\label{eq:Qs}\begin{split}
Q_1(x,t)  &= \frac{1}{2}  \left(\begin{array}{cc} 0 & \q \\ -\p & 0  \end{array}\right), \quad
Q_2(x,t)  = \frac{1}{2}  \left(\begin{array}{cc} \r & 0 \\ 0 & \s   \end{array}\right), \quad
Q_3(x,t) = \frac{1}{2}  \left(\begin{array}{cc} 0 & \v \\ -\w & 0  \end{array}\right).
\end{split}\end{equation}
With such choice for $\xi(x,t,\lambda)$ we obviously have
\begin{equation}\label{eq:xi1}\begin{split}
 \lim_{\lambda\to\infty} \xi(x,t,\lambda) = \openone
\end{split}\end{equation}
which provides the canonical normalization of the RHP. Besides we have requested that $\mathcal{Q}(x,t,\lambda)$
takes values in the Kac-Moody algebra determined by the grading (\ref{eq:g0}); in other words $\mathcal{Q}(x,t,\lambda)$
satisfies
\begin{equation}\label{eq:Q3}\begin{split}
 \mathcal{Q}(x,t,\lambda) = C_0 \mathcal{Q}(x,t,-\lambda) C_0^{-1}, \qquad C_0= \exp(\pi i J).
\end{split}\end{equation}
Then we can introduce $U(x,t,\lambda)$ and $V(x,t,\lambda)$ as the non-negative parts of \cite{GeDik,DriSok}:
\begin{equation}\label{eq:UV1}\begin{split}
 U(x,t,\lambda) = -\left( \lambda^a \xi (x,t,\lambda) J  \xi^{-1} (x,t,\lambda) \right)_+, \qquad
 V(x,t,\lambda) = -\left( \lambda^b \xi (x,t,\lambda) J  \xi^{-1} (x,t,\lambda) \right)_+,
\end{split}\end{equation}
where $a$ and $b$ can be any integers. For simplicity and definiteness we will fix up $a=2$ and $b=4$.
The explicit calculation of $U(x,t,\lambda)$ and $V(x,t,\lambda)$ in terms of $Q_s(x,t)$ can be done using
the well known formula
\begin{equation}\label{eq:UV3}\begin{split}
\xi (x,t,\lambda) J  \xi^{-1} (x,t,\lambda) = J + \sum_{s=1}^{\infty} \frac{1}{s!} \ad_{\mathcal{Q}}^s J, \qquad
\ad_{\mathcal{Q}} J = [\mathcal{Q}, J], \qquad    \ad_{\mathcal{Q}}^2 J =  [\mathcal{Q},[\mathcal{Q}, J]], \quad  \dots
\end{split}\end{equation}
Since in (\ref{eq:UV1}) we need only the non-negative powers of $\lambda$ for any $a$ and $b$ we will need
only finite number of terms. In particular for $a=2$ and $b=4$ we have:

\begin{equation}\label{eq:1aU}\begin{split}
U(x,t,\lambda) &=- \left( \lambda^2 \xi J\hat{\xi}\right)_+ = -\lambda^2 J + \lambda Q(x,t) + U_2(x,t) , \\
Q(x,t) &= - [Q_1,J] = \left(\begin{array}{cc} 0 & \q \\ \p & 0  \end{array}\right), \\
U_2(x,t) &= - \frac{1}{2} [Q_1, [Q_1, J]] - [Q_2(x,t),J] = \frac{1}{2} \left(\begin{array}{cc} \q \p & 0 \\ 0 & -\p \q \end{array}\right).
\end{split}\end{equation}
Note that since $Q_2(x,t)\in \mathfrak{g}^{(0)}$ then $[Q_2(x,t),J]=0$. Similarly
\begin{equation}\label{eq:V1-4}\begin{split}
 V(x,t,\lambda) &= -\left( \lambda^4 \xi^\pm J \hat{\xi}^\pm (x,t,\lambda)\right)_+ =
 V_4(x,t) +\lambda V_3(x,t) +\lambda^2 V_2(x,t) +\lambda^3 Q(x,t)  - \lambda^4 J, \\
V_2(x,t) &=  U_2(x,t), \qquad V_3(x,t) = -\frac{1}{2}\ad_{Q_2}\ad_{Q_1}J - \frac{1}{6}\ad_{Q_1}^3J    , \\
V_4(x,t) &=  -\frac{1}{2} \left(\ad_{Q_3}\ad_{Q_1}J + \ad_{Q_1}\ad_{Q_3}J\right) - \frac{1}{6} \left(
\ad_{Q_1}\ad_{Q_2}\ad_{Q_1}J+ \ad_{Q_2} \ad_{Q_1}^2J \right) - \frac{1}{24} \ad_{Q_1}^4J.
\end{split}\end{equation}
Here we used again $[Q_2(x,t),J]=0$ and $[Q_4(x,t),J]=0$. Below we will pay special attention to the particular case $p=1$ which
corresponds to the vector GI equation.

\subsection{ RHP and multi-component GI equations}

Here we assume that the FAS of $L$ and $M$ satisfy a canonical RHP with special reduction:
\begin{equation}\label{eq:xi00}\begin{split}
\xi^\pm(x,t, -\lambda) = \xi^{\pm, -1}(x,t,\lambda) ,
\end{split}\end{equation}
i.e., $\mathcal{Q}(x,t,\lambda) = -\mathcal{Q}(x,t,-\lambda)$ and therefore $Q_{2s}(x,t)=0$.
 As a result the expression for the Lax pair simplifies to
 \begin{equation}\label{eq:LM1}\begin{aligned}
L\psi &\equiv i \frac{\partial \psi}{ \partial x } +U(x,t,\lambda)\psi=0,  &\quad M\psi  &\equiv i \frac{\partial \psi}{ \partial t } +V(x,t,\lambda)\psi=0,\\
 U(x,t,\lambda) &= U_2(x,t) +\lambda Q(x,t) - \lambda^2 J, &\quad  Q(x,t) &= \left(\begin{array}{cc} 0 & \q \\ \p & 0  \end{array}\right),
 \qquad U_2(x,t) =  \frac{1}{2}\left(\begin{array}{cc} \q \p & 0 \\ 0 & -\p \q \end{array}\right), \\
V_2(x,t) &= U_2(x,t), &\quad V_3(x,t) &= \left(\begin{array}{cc} 0 & \v -\frac{1}{6}\q \p \q \\ \w - \frac{1}{6} \p \q \p  \end{array}\right),
 \end{aligned}\end{equation}
 \[
 V_4(x,t) =  \frac{1}{2} \left(\begin{array}{cc} \q \w+ \v \p -\frac{1}{12} \q \p \q \p & 0 \\ 0 & - \w \q- \p \v + \frac{1}{12} \p \q \p \q  \end{array}\right).
 \]
The commutation  $[L,M]$ must vanish identically with respect to $\lambda$. It is polynomial in $\lambda$ with the
following coefficients:
\begin{equation}\label{eq:LM1'}\begin{aligned}
& \lambda^5: &\quad & -[J,V_1] - [Q,J] =0, &\quad  &\Rightarrow &\quad V_1&=Q, \\
& \lambda^4: &\quad & -[J,V_2] + [Q,V_1] - [U_2,J] =0, &\quad &\Rightarrow &\quad &\mbox{identity} \\
& \lambda^3: &\quad & i \frac{\partial V_1}{ \partial x } +[U_2,V_1] + [Q,V_2] =[J, V_3],
\end{aligned}\end{equation}
The last of these equations is fulfilled iff
\begin{equation}\label{eq:v-w}\begin{split}
\v= \frac{i}{2} \frac{\partial \q}{ \partial x } + \frac{1}{6} \q \p \q    , \qquad \w= -\frac{i}{2} \frac{\partial \p}{ \partial x } + \frac{1}{6} \p \q \p.
\end{split}\end{equation}
The next equations are:
\begin{equation}\label{eq:NLEE1}\begin{aligned}
& \lambda^2: &\quad & i \frac{\partial V_2}{ \partial x } +[U_2,V_2] + [Q,V_3] =[J, V_4]\equiv 0,  \\
& \lambda^1: &\quad & i \frac{\partial V_3}{ \partial x } - i\frac{\partial Q}{ \partial t } +[U_2,V_3] + [Q,V_4] = 0,  &\quad
 \lambda^0: &\quad & i \frac{\partial V_4}{ \partial x } - i\frac{\partial U_2}{ \partial t}+ [U_2,V_4] = 0.
\end{aligned}\end{equation}
The first of the above equations is satisfied identically. The second one written in block-components gives the following NLEE which can be viewed as
multicomponent GI equations related to the {\bf D.III} symmetric space:
\begin{equation}\label{eq:NLEE2}\begin{split}
i \frac{\partial \q}{ \partial t} + \frac{1}{2} \frac{\partial^2 \q }{ \partial x^2 } - \frac{i}{2} \q \frac{\partial \p}{ \partial x } \q + \frac{1}{4} \q \p \q \p \q &=0, \\
-i \frac{\partial \p}{ \partial t} + \frac{1}{2} \frac{\partial^2 \p }{ \partial x^2 } + \frac{i}{2} \p \frac{\partial \q}{ \partial x } \p + \frac{1}{4} \p \q \p \q \p &=0.
\end{split}\end{equation}
The last equation in (\ref{eq:NLEE1}) is a consequence of  the expressions for $Q_0$ and $V_4$ and the
eqs. (\ref{eq:NLEE2}) and  (\ref{eq:v-w}).

\subsection{ RHP and multi-component Kaup-Newel  equations}

The Lax pair for the KN-system is obtained from  (\ref{eq:LM1}) by a gauge transformation:
\begin{equation}\label{eq:ga0}\begin{split}
\tilde{L} \tilde{\chi}^\pm(x,t,\lambda) &\equiv i \frac{\partial \tilde{\chi}^\pm}{ \partial x } + (\lambda \tilde{Q}(x,t) -\lambda^2 J)
 \tilde{\chi}^\pm(x,t,\lambda)=0, \\
\tilde{M} \tilde{\chi}^\pm(x,t,\lambda) &\equiv i \frac{\partial \tilde{\chi}^\pm}{ \partial t } + (\lambda \tilde{V}_3(x,t) + \lambda^2 \tilde{V}_2 (x,t)
+ \lambda^3 \tilde{Q}(x,t) -\lambda^4 J)  \tilde{\chi}^\pm(x,t,\lambda)=0,
\end{split}\end{equation}
where
\begin{equation}\label{eq:Qt}\begin{split}
\tilde{V}_3(x,t) = g_0^{-1} V_3 (x,t)g_0(x,t) , \qquad \tilde{V}_2(x,t) = g_0^{-1} V_2 (x,t)g_0(x,t) ,
\qquad \tilde{Q}(x,t) = g_0^{-1} Q (x,t)g_0(x,t)
\end{split}\end{equation}
and the gauge $g_0(x,t)$ is defined uniquely by the equations:
\begin{equation}\label{eq:ga1}\begin{split}
 i \frac{\partial g_0}{ \partial x } +U_2(x,t) g_0 (x,t) =0, \qquad  i \frac{\partial g_0}{ \partial t } +V_4(x,t) g_0 (x,t) =0.
\end{split}\end{equation}
Note that $g_0(x,t)$ must be block-diagonal, so similarity transformations with it preserve the grading
(the block-matrix structure) of the coefficients in $U(x,t,\lambda)$ and $V(x,t,\lambda)$; in particular,
$g_0^{-1}J g_0(x,t) =J$. Therefore we introduce
\begin{equation}\label{eq:Qt'}\begin{split}
 \tilde{Q} = \left(\begin{array}{cc} 0 & \tilde{\q} \\ \tilde{\p} & 0   \end{array}\right)
\end{split}\end{equation}
Applying the gauge transformation to $V_2(x,t)$ we easily obtain:
\begin{equation}\label{eq:V2}\begin{split}
\tilde{V}_2(x,t) =  - \frac{1}{2} g_0^{-1}  [Q_1, [Q_1(x,t),J]] = \frac{1}{2} \left(\begin{array}{cc} \tilde{\q}\tilde{\p}
& 0 \\ 0 & - \tilde{\p}\tilde{\q} \end{array}\right).
\end{split}\end{equation}
Next we apply the compatibility condition of $\tilde{L}$ and $\tilde{M}$ and obtain
\begin{equation}\label{eq:V3}\begin{split}
V_3 = \ad_J^{-1} \left( i \frac{\partial \tilde{Q}}{ \partial x }  + [\tilde{Q}, V_2]\right) = \frac{1}{2} \left(\begin{array}{cc}
0 & i \tilde{\q}_x - \tilde{\q}\tilde{\p}\tilde{\q} \\ - i \tilde{\p}_x + \tilde{\p}\tilde{\q}\tilde{\p} \end{array}\right).
\end{split}\end{equation}
Finally the multi-component Kaup-Newell equation takes the form:
\begin{equation}\label{eq:KNm}\begin{split}
i \frac{\partial \tilde{\q}}{ \partial t } &+ \frac{\partial^2 \tilde{\q} }{ \partial x^2 } + i \frac{\partial \tilde{\q}\tilde{\p}\tilde{\q}}{ \partial x }=0, \\
-i \frac{\partial \tilde{\p}}{ \partial t } &+ \frac{\partial^2 \tilde{\p} }{ \partial x^2 } - i \frac{\partial \tilde{\p}\tilde{\q}\tilde{\p}}{ \partial x }=0.
\end{split}\end{equation}

\section{Effects of reductions on the scattering matrix}
Let us briefly outline the effects of the reductions on the scattering matrix. We will start
with the properties of the  Jost solutions and then  with determining the reduction of the scattering data.
In particular we will find the symmetries on the discrete eigenvalues of $L$.

\subsection{The local involution case}
Each of the reductions in (\ref{eq:Z2-Mi}) has a natural action on the Jost solutions and, as a consequence
on the FAS and the scattering matrix. In particular the reduction 1) of eq. (\ref{eq:Z2-Mi}) requires that:
\begin{equation}\label{eq:locR2'}\begin{split}
A_1 \psi ^\dag (x,t,\lambda^*)A_1^{-1} = \psi^{-1}  (x,t,\lambda) , \qquad  A_1 \phi ^\dag (x,t,\lambda^*)A_1^{-1} = \phi^{-1}  (x,t,\lambda) , \\
A_1 T ^\dag (t,\lambda^*)A_1^{-1} = T^{-1}  (t,\lambda) ,
\end{split}\end{equation}
where $A_1$ was introduced in eq. (\ref{eq:A13}). So for the matrix elements of the scattering matrix we have
\begin{equation}\label{eq:Tdag2}\begin{aligned}
a_1 (a^+(\kappa_1\lambda^*))^\dag a_1^{-1}  & = c^-(\lambda), & \qquad a_2 (a^-(\kappa_1\lambda^*))^\dag a_2^{-1}  & = c^+(\lambda), \\
a_1 (b^+(\kappa_1\lambda^*))^\dag a_2^{-1}  & = d^-(\lambda), & \qquad a_2 (b^-(\kappa_1\lambda^*))^\dag a_1^{-1}  & = d^+(\lambda),
\end{aligned}\end{equation}
If we specify $a_1=\openone$ and $a_2 = \epsilon_2 \openone$, $\epsilon_2 =\pm 1$ we get:
\begin{equation}\label{eq:ab*1}\begin{aligned}
(a^\pm (\kappa_1\lambda^*))^\dag & = c^\mp (\lambda), & \qquad  \epsilon_2 (b^\pm (\kappa_1\lambda^*))^\dag & = d^\mp (\lambda),\\
(\rho^+ (\kappa_1\lambda^*))^\dag & = \epsilon_2 \rho^- (\lambda), & \qquad (\tau^+ (\kappa_1\lambda^*))^\dag & = \epsilon_2 \tau^- (\lambda).
\end{aligned}\end{equation}
It is well known that the zeros of $a^\pm(\lambda)$ and  $c^\pm(\lambda)$  are the discrete eigenvalues of $L$. From (\ref{eq:ab*1}) one finds that
if $\lambda_1^+ \in \Omega_1\cup \Omega_3$ is a zero of $a^+(\lambda)$ then $\kappa_1(\lambda_1^+)^* \in \Omega_2\cup \Omega_4$
is a zero of $c^-(\lambda)$.

The other local reduction 3) in (\ref{eq:Z2-Mi}) is treated similarly. Here for simplicity we assume $p=q$, i.e. the matrices
$\p(x,t)$ and $\q(x,t)$ are square. Then
\begin{equation}\label{eq:locR3}\begin{split}
A_3 \psi ^* (x,t,\lambda^*)A_3^{-1} = \psi  (x,t,\lambda) , \qquad  A_3 \phi ^\dag (x,t,\lambda^*)A_3^{-1} = \phi (x,t,\lambda) , \qquad
A_3 T ^* (t,\lambda^*)A_3^{-1} = T  (t,\lambda) ,
\end{split}\end{equation}
with  $A_3$ defined by (\ref{eq:A13}), which means that
\begin{equation}\label{eq:Tdag3}\begin{aligned}
b_2 (a^+(\kappa_1\lambda^*)^* b_2^{-1}  & = a^-(\lambda), & \qquad b_1 (a^-(\kappa_1\lambda^*)^* b_1^{-1}  & = a^+(\lambda), \\
b_1 (b^+(\kappa_1\lambda^*)^* b_1^{-1}  & = -b^-(\lambda), & \qquad b_2 (b^-(\kappa_1\lambda^*)^* b_1^{-1}  & = -b^+(\lambda),
\end{aligned}\end{equation}
If we assume $b_1=\openone$ and $b_2 = \epsilon_2 \openone$, $\epsilon_2 =\pm 1$ we get:
\begin{equation}\label{eq:ab*2}\begin{aligned}
(a^\pm (\kappa_1\lambda^*))^* & = a^\mp (\lambda), & \qquad  \epsilon_2 (b^\pm (\kappa_1\lambda^*))^* & = -b^\mp (\lambda),\\
(\rho^+ (\kappa_1\lambda^*))^* & = -\epsilon_2 \rho^- (\lambda), & \qquad (\tau^+ (\kappa_1\lambda^*))^* & = -\epsilon_2 \tau^- (\lambda).
\end{aligned}\end{equation}

\subsection{The nonlocal involution case}

These reductions also have a natural (but different from the above) effect on the
Jost solutions and the scattering matrix. The reduction 1) from (\ref{eq:Z2-nl}) leads to:
\begin{equation*}\label{eq:nlocR2'}\begin{split}
C_1 \phi ^\dag (x,t, \kappa_1\lambda^*)C_1^{-1} &= \psi^{-1}  (-x,-t,-\lambda) , \qquad
C_1 \psi ^\dag (x,t, \kappa_1\lambda^*)C_1^{-1} = \phi^{-1}  (-x,-t,-\lambda) , \\
C_1 T ^\dag (t,\kappa_1\lambda^*)C_1^{-1} &= T  (-t,\lambda) ,
\end{split}\end{equation*}
where $C_1$ is given by (\ref{eq:C13}).
As a consequence for the matrix elements of $T(t,\lambda)$ we get:
\begin{equation}\label{eq:Tnl}\begin{aligned}
 c_1 (a^-(\kappa_1\lambda^*))^\dag c_1^{-1} & = a^+(\lambda), &\qquad  c_1 (b^-(\kappa_1\lambda^*))^\dag c_2^{-1} & = b^-(\lambda),
 \qquad  c_2 (b^+(\kappa_1\lambda^*))^\dag c_1^{-1} & = b^+(\lambda),
\end{aligned}\end{equation}
 The reduction 3) from (\ref{eq:Z2-nl}) leads to:
\begin{equation}\label{eq:nlocR2''}\begin{split}
C_3  \phi ^* (x,t, \kappa_1\lambda^*)C_3 ^{-1} &= \psi^{-1}  (-x,-t,\lambda) , \qquad
C_3  \psi ^* (x,t, \kappa_1\lambda^*)C_3 ^{-1} = \phi^{-1}  (-x,-t,\lambda) , \\
C_3  T ^* (t,\kappa_1\lambda^*)C_3 ^{-1} &= T ^{-1} (-t,\lambda) ,
\end{split}\end{equation}
with $C_3$ defined by (\ref{eq:C13}).
Thus for the matrix elements of $T(t,\lambda)$ we find:
\begin{equation}\label{eq:Tnl2}\begin{aligned}
d_1 (a^+(\kappa_1\lambda^*))^* d_1^{-1} & = c^-(\lambda), &\qquad  d_2 (a^-(\kappa_1\lambda^*))^* d_2^{-1} & = c^+(\lambda),\\
d_2 (b^+(\kappa_1\lambda^*))^* d_1^{-1} & = -d^+(\lambda), &\qquad  d_1 (b^-(\kappa_1\lambda^*))^* d_2^{-1} & = -d^-(\lambda),
\end{aligned}\end{equation}
Particularly if we put $d_1=\openone$ and $d_2=\epsilon_2 \openone$ we find:
\begin{equation}\label{eq:3*1}\begin{aligned}
(a^\pm(\kappa_1\lambda^*))^* & = c^\mp (\lambda), &\qquad (b^\pm(-t,\kappa_1\lambda^*))^*=-\epsilon_2 d^\pm(\lambda),
\end{aligned}\end{equation}
and
\begin{equation}\label{eq:3*2}\begin{split}
 \rho^\pm(-t,\kappa_1\lambda^*) = -\epsilon_2 \tau^\mp (\lambda).
\end{split}\end{equation}
From (\ref{eq:3*1}) we see that if $\lambda_1^+ \in \Omega_1\cup \Omega_3$ is a zero of $a^+(\lambda)$ then
$\kappa_1(\lambda_1^+)^* \in \Omega_2\cup \Omega_4$ is a zero of $c^-(\lambda)$.

\section{Soliton solutions}
Here we adapt Zakharov-Shabat's dressing method \cite{ZaSh*74a,ZaSh*79} to the above Lax pairs.

\subsection{Dressing method}
One of the most convenient approaches to the derivation of the soliton solutions is the so-called dressing method \cite{ZaSh*74a,ZaSh*79} (see also \cite{67,GVYa*08,MiZa*80,RI,TV2,GGK*XX}).
The rationale of the method is the construction of a nontrivial (dressed) FAS, $\chi^{\pm}(x,t,\lambda)$ from the known (bare) FAS,
$\chi_0^{\pm}(x,t,\lambda)$ by the means of the so-called dressing factor $u(x,t,\lambda):$
\begin{equation}\label{eq:chi}\begin{split}
 \chi^{\pm}(x,t,\lambda)=u(x,t,\lambda)\chi_0^{\pm}(x,t,\lambda).
\end{split}\end{equation}
The dressing factor is analytic in the entire complex $\lambda$-plane, with the exception of the newly added  simple pole
singularities at  $\lambda=\lambda_k^{\pm},$  $k=1,2, \ldots, N:$. It is known that these singularities are in fact discrete eigenvalues
of the `dressed' Lax operator $L$:
 \begin{equation} \label{eq:u}
u(x,t,\lambda)= \openone + \sum_{k=1} ^{N}\left ( \frac{\lambda_1^+- \lambda_1^{-}}{\lambda- \lambda_1^{+}}
B_k(x,t) + \frac{\lambda_1^-  -  \lambda_1^{+}}{\lambda- \lambda_1^{-}} \tilde{B}_k(x,t) \right).
\end{equation}
  As far as the FAS satisfy the Lax pair equations \eqref{eq:LM1}, the dressing factor must be a solution of the equation
\begin{equation} \label{dress2}
 i u_x +U_2 u -u U_2^{(0)}+ \lambda(Q u - u Q^{(0)})+\lambda^2 [u,J] =0,
\end{equation} where the upper index $(0)$ indicates the quantities, associated to the bare solution. The equation \eqref{dress2} must
 hold identically with respect to $\lambda.$  Since $u$ has poles at finitely many points of the discrete spectrum, it will be enough to
  request that \eqref{dress2} holds for $\lambda \to \infty$ and $\lambda \to \lambda_{k}^{\pm}.$ For $\lambda \to \infty,$  $u \to \openone,$
   so the derivative term in \eqref{dress2} disappears. The $\lambda^2- $ terms are proportional to $[J, \openone]$ that also identically
  vanishes. Thus, we are left with two terms, which are easily evaluated to be
\begin{equation}\label{eq:QP}\begin{aligned}
&\lambda^1: & Q-Q^{(0)} &= \sum_{k=1}^{N}(\lambda_{k}^+-\lambda_k^-)[J,B_k-\tilde{B}_k], \\
&\lambda^0: & U_2-U_2^{(0)}&= \sum_{k=1}^{N}(\lambda_{k}^+-\lambda_k^-)\left([J,\lambda_k^+B_k-\lambda_k^-\tilde{B}_k]
-Q(B_k-\tilde{B}_k) +(B_k-\tilde{B}_k)Q^{(0)} \right) .
\end{aligned}\end{equation}
Thus, if we know the residues $B_k, \tilde{ B}_k$ we are able to reconstruct $Q(x,t)$ and $U_2(x,t).$  The condition that
\eqref{dress2} holds for $\lambda \to \lambda_k^{\pm}$ leads to the following:
\begin{equation} \label{eq:Bk}
\begin{split}
 i \partial_x B_k +(U_2 +\lambda_k^+ Q) B_k -B_k(U_2^{(0)}+\lambda_k^+  Q^{(0)})+(\lambda_k^+)^2 [B_k,J] =0.
\end{split}
\end{equation}
In the simplest possible nontrivial case, $B_k$ are rank 1 matrices of the form \begin{equation} \label{eq:rankone}
B_k=|n_k\rangle \langle m_k|
\end{equation} satisfying the matrix equation \eqref{eq:Bk}, ($|n \rangle$ is a vector-column, $ \langle m| $ is a vector-row as usual).
It is straightforward to verify that $B_k$ in the form \eqref{eq:rankone} will satisfy \eqref{eq:Bk}, if and only if
\begin{equation} \label{eq:mn}
\begin{split}
i\partial_x |n_k \rangle  + \left(U_2^{(0)} + \lambda_k^+ Q^{(0)} - (\lambda_k^+)^2 J\right )|n_k \rangle &=0, \\
i\partial_x \langle m_k| - \langle m_k| \left(U_2^{(0)} + \lambda_k^+ Q^{(0)} - (\lambda_k^+)^2 J\right)&=0,
\end{split}
\end{equation}    i.e.
\begin{equation} \label{eq:mn2}
|n_k \rangle=\chi^{+}(x,t,\lambda_k^+)|n_{k,0} \rangle, \qquad \langle m_k|=\langle m_{k,0}|\hat{\chi}_0^{+}(x,t,\lambda_k^+),
\end{equation}
where $|n_{k,0} \rangle$ and $\langle m_{k,0}|$ are some constant vectors. One can start with the trivial bare solutions
 $Q^{(0)}=0,$ $U_2^{(0)}=0,$ so that $\chi^{+}_0(x,t,\lambda)=\exp i(\lambda^2 J x +\lambda^4 J t)$ is known explicitly.

\subsection{Example - One soliton solution with local reduction}

In the first example the dressing factor $u(x,\lambda;t)$   satisfies the reduction conditions from the first reduction of \eqref{eq:Z2-Mi}:

\begin{equation}\label{eq:g_reduction}\begin{aligned}
&\mbox{A)} & \quad A_1 u^\dag (x,t,\kappa_1\lambda^*) A_1^{-1} &= u^{-1}(x,t,\lambda), \qquad
\mbox{B)} & \quad u (x,t,-\lambda)  &= u^{-1}(x,t,\lambda) .
\end{aligned}\end{equation}
We consider the case $p=1,$ i.e. $\q$ is a vector-row and $\p$ is a vector-column, $J$ is diagonal with
$J_{11}=1$ and $J_{ii}=-1$ for $i=2, \ldots n.$  $(A_1)_{ij}=\epsilon_i \delta_{ij}$ is diagonal, with $\epsilon_i=\pm1.$
Introducing the notation
\[ A_1=\text{diag}(a_1,a_2) \]
 for the block-diagonal matrix $A_1$   and noting that $A_1=A_1^{-1},$ we have the following relations between $\p$ and $\q$:
\begin{equation}\label{eq:g_reduction'}\begin{split}
\q=  \kappa_1 a_1 \p ^{\dag}a_2,  \qquad  \p=  \kappa_1 a_2 \q ^{\dag}a_1.
\end{split}\end{equation}
A dressing factor with simple poles at $\lambda=\lambda_1^{\pm}$ has the form
  \begin{equation} \label{dress1p}
 u(x,t,\lambda)= \openone+\frac{\lambda_1^+-\lambda_1^-}{\lambda-\lambda_1^+}B_1(x,t)
 +\frac{\lambda_1^--\lambda_1^+}{\lambda-\lambda_1^-}\tilde{B}_1(x,t)
\end{equation}
Moreover, the reductions $A$ and $B$ are simultaneously satisfied if
 \begin{equation} \label{reduct_comp}
 \lambda_1^+= -\kappa_1 (\lambda_1^-)^*, \qquad \tilde{B}_1=A_1B_1^{\dag} A_1^{-1}.
\end{equation}
Let us introduce the notation $\mu\equiv \lambda_1^+$ and in polar form $\mu=\rho e^{i\varphi}.$
Both reductions $A,B$ must hold identically with respect to $\lambda$ which necessitates (e.g. when $\lambda \to \mu$ )
  \begin{equation} \label{B_1}
B_1 \left( \openone-\frac{\mu+ \kappa_1 \mu^*}{2\mu}B_1(x,t)
 +\frac{\mu + \kappa_1 \mu^*}{\mu-\kappa_1\mu^*}A_1B_1^{\dag}(x,t)A_1^{-1} \right)=0
\end{equation}
Looking for a rank one solution $B_1=|n\rangle \langle m| $ of the matrix equation \eqref{B_1},
($|n \rangle$ is a vector-column, $ \langle m| $ is a vector-row as usual) we find that
  \begin{equation} \label{B_1sol}
B_1 = z \frac{A_1 |m^* \rangle \langle m|}{\langle m |A_1 |m^* \rangle}
\end{equation} where the complex constant $z$ satisfies the linear equation
 \begin{equation} \label{z_eq}
1-\frac{\mu+ \kappa_1 \mu^*}{2\mu}z
 +\frac{\mu + \kappa_1 \mu^*}{\mu-\kappa_1\mu^*}z^*=0.
\end{equation}
In addition, from \eqref{eq:QP}-- \eqref{eq:Bk} it follows that
\begin{equation} \label{dress1a}
\begin{split}
& i \partial_x B_1 +(U_2 +\mu Q) B_1 -B_1(U_2^{(0)}+\mu  Q^{(0)})+\mu^2 [B_1,J] =0,\\
& Q=Q^{(0)}+(\mu+ \kappa_1 \mu^*)[J, B_1 - C_1B_1^{\dag}(x,t)C_1^{-1}].
\end{split}
\end{equation}
and together with the assumption $B_1=|n\rangle \langle m| $ one can find out that $\langle m| $  satisfies the bare equation
\begin{equation} \label{m}
i\partial_x \langle m| - \langle m| (U_2^{(0)} + \mu Q^{(0)} - \mu^2 J)=0. \end{equation} Therefore, starting from the
trivial solution $U_2^{(0)} = Q^{(0)}=0$ we find
\begin{equation} \label{sol_m}
 \langle m|= \langle m_0|e^{i(\mu^2 x + \mu^4 t)J},
\end{equation}
where $\langle m_0|$ is a constant vector with components $m_{0j}.$ Now we can write the one-soliton solution,
\begin{equation} \label{1sol}
 \q_{j-1}(x,t)=Q_{1j}=4 \rho r(\kappa_1) \frac{m_{0j}e^{\xi_0}e^{-i\phi(x,t)}}{m_{01}\cosh(\theta(x,t)-\xi_0)}, \qquad j=2,\ldots,n,
\end{equation}
where $r(1)=i\sin \varphi$, and $r(-1)= \cos \varphi$ and when $A_1=\openone,$
\[ e^{-2\xi_0}\equiv \frac{\sum_{j=2}^n  |m_{0j}|^2}{|m_{01}|^2} \]
is real and positive,
\begin{equation} \label{Den}\begin{split}
\theta(x,t)&=2\rho^2(\sin 2 \varphi) x + 2 \rho^4 (\sin 4 \varphi) t ,\qquad \phi(x,t) =2\rho^2(\cos 2\varphi) x + 2 \rho^4 (\cos 4\varphi) t,
\end{split}\end{equation}

\subsection{Example - One soliton solution with nonlocal reduction}

 In the second example the dressing factor $u(x,t,\lambda;t)$  satisfies the reduction conditions from the first reduction of \eqref{eq:Z2-nl}:
\begin{equation}\label{eq:Z2-nl2}\begin{split}
\mbox{A)}  \quad C_1 u^\dag (-x,-t,\kappa_1\lambda^*) C_1^{-1} = u^{-1}(x,\lambda), \qquad
\mbox{B)}  \quad u (x,t,-\lambda) = u^{-1}(x,\lambda) .
\end{split}\end{equation}
Let us take for simplicity $p=1,$ $n=2,$ $\p$ and $\q$ are scalar functions. The automorphism $C_1$ cannot be
represented by a diagonal matrix, since now it must change the sign of $J\equiv \sigma_3$. Hence, we take
\begin{equation}\label{eq:C_1nl}\begin{split}
 C_1 = \left(\begin{array}{cc} 0 & 1 \\ 1 & 0   \end{array}\right).
\end{split}\end{equation}
 The reduction gives now the following connections between $\p$ and $\q$, under which the equations are $\mathcal{CPT}$-invariant:
\begin{equation}\label{eq:nlpq_reduction}\begin{split}
\q(x,t)=  -\kappa_1 \q ^{*}(-x,-t),  \qquad \p(x,t)=  -\kappa_1  \p ^{*}(-x,-t) .
\end{split}\end{equation}
The dressing factor satisfies the equation \eqref{dress2}. Again it is taken to have simple poles at $\lambda=\lambda_1^{\pm}:$
  \begin{equation} \label{dress1p_nl}
 u(x,t,\lambda)= \openone+\frac{\lambda_1^+-\lambda_1^-}{\lambda-\lambda_1^+}B_1(x,t)
 +\frac{\lambda_1^--\lambda_1^+}{\lambda-\lambda_1^-}\tilde{B}_1(x,t)
\end{equation}
This time the reductions $A$ and $B$ are simultaneously satisfied if
 \begin{equation} \label{reduct_comp_nl}
 \lambda_1^+= -\kappa_1 (\lambda_1^-)^*, \qquad \tilde{B}_1(x,t)=C_1B_1^{\dag}(-x,-t) C_1^{-1}.
\end{equation}
With the short notations $\mu\equiv \lambda_1^+=\rho e^{i\varphi}$ we obtain the equation for $B_1(x,t)$
  \begin{equation} \label{B_1nl}
B_1(x,t) \left( \openone-\frac{\mu+ \kappa_1 \mu^*}{2\mu}B_1(x,t)
 +\frac{\mu + \kappa_1 \mu^*}{\mu-\kappa_1\mu^*}C_1B_1^{\dag}(-x,-t)C_1^{-1} \right)=0
\end{equation} with a rank one solution  \begin{equation} \label{B_1solnl}
B_1 (x,t)= z \frac{C_1 |m^*(-x,-t) \rangle \langle m(x,t)|}{\langle m(x,t) |C_1 |m^*(-x,-t) \rangle}
\end{equation} where the complex constant $z$ and the components of $\langle m(x,t)|,$ i.e. $m_j(x,t)$ are as before. The solution is \begin{equation} \label{sol_nl}
Q(x,t)=Q^{(0)}(x,t)+(\mu+ \kappa_1 \mu^*)[J, B_1(x,t) - C_1B_1^{\dag}(-x,-t)C_1^{-1}].
\end{equation}
Starting with $Q^{(0)}(x,t)\equiv 0$ and real $m_{0j}$ we obtain
\begin{equation} \label{1solnl} \begin{split}
 \q(x,t)&=Q_{12}=(\mu+\kappa_1 \mu^*)(z-z^*) \frac{m_{02}e^{-i\phi(x,t)}}{m_{01}\cosh(\theta(x,t))}, \\
\p(x,t)&=Q_{21}=-(\mu+\kappa_1 \mu^*)(z-z^*)\frac{m_{01}e^{i\phi(x,t)}}{m_{02}\cosh(\theta(x,t))},
\end{split}
\end{equation}
with $\phi(x,t)$ and $\theta(x,t)$ defined as before, $$(z-z^*)_{\kappa_1=1}=2i\tan \varphi,  \qquad (z-z^*)_{\kappa_1=-1}=-2i\cot \varphi. $$
Note that for the solutions \eqref{1solnl} the property \eqref{eq:nlpq_reduction} is evident.
It is worth noting that in both cases the action of the reduction on $\lambda$ is $\lambda \to \epsilon \lambda^*$.
In both cases the action on $\lambda$ is very nice. Indeed, the analyticity regions are $A_+=\im \lambda^2 >0$
and $A_-=\im \lambda^2 <0$. The action on $\lambda$ always maps $A_+\to A_-$.

\section{Integrals of motion of the multi-component DNLS equations}

From eq. (\ref{eq:5.2}) we conclude that block-diagonal Gauss factors $D_J^\pm(\lambda)$ are generating functionals of the
integrals of motion. The principal series of integrals is generated by $m_1^\pm(\lambda)$:
\begin{equation}\label{eq:m1pm}
\pm \ln m_1^\pm = \frac{1}{i} \sum_{s=1}^\infty I_s \lambda^{-s}.
\end{equation}
Let us first outline a way to calculate their densities as functionals of $Q(x,t)$. To do this we make use of the third type
of Wronskian identities involving $\dot{\chi}^\pm(x,\lambda)$. They have the form:
\begin{equation}\label{eqB:wr.dd1'}\begin{split}
& \left.  \left( i\hat{\xi }^\pm \dot{\xi}^\pm (x,\lambda )  +2\lambda Jx \right) \right|_{x=-\infty }^{\infty } =  -\int_{-\infty
}^{\infty } d x\, \left( \hat{\xi }    (Q(x) -2\lambda  J)\xi (x,\lambda ) + \lambda^2[ J, \hat{\xi}^\pm \dot{\xi}^\pm(x,\lambda)]\right),
\end{split}\end{equation}
If we multiply both sides of (\ref{eqB:wr.dd1'}) with $J$ and take the Killing form we get:
\begin{equation}\label{eqB:wr.d1'}\begin{split}
& \left.  \left\langle ( i\hat{\xi }^\pm \dot{\xi}^\pm (x,\lambda) +2  \lambda Jx, J \right\rangle \right|_{x=-\infty }^{\infty } = \pm
2i \frac{d}{d\lambda} \ln m_1^\pm (\lambda),
\end{split}\end{equation}
which means that
\begin{equation}\label{eq:m1pm'}\begin{split}
\pm i \frac{d}{d\lambda} \ln m_1^\pm (\lambda)= \frac{i}{2} \int_{-\infty}^{\infty } d x\, \left( \langle
(Q(x) - 2\lambda J) ,\xi^\pm (x,\lambda) J\hat{\xi}^\pm (x,\lambda ) \rangle +2\lambda \langle J, J\rangle \right).
\end{split}\end{equation}
If we introduce the notations:
\begin{equation}\label{eq:xiX}\begin{split}
\xi^\pm J \hat{\xi}^\pm (x,\lambda)  = J + \sum_{s=1}^{\infty} \lambda^{-s} X_s,
\end{split}\end{equation}
then from eq. (\ref{eq:UV3}) one can calculate recursively $X_s$. Of course their complexity grows
rather quickly with $s$. Knowing $X_s$ we find the following recursive formula for $I_s$:
\begin{equation}\label{eq:Is1}\begin{split}
I_{2s} = \frac{1}{4s} \int_{-\infty}^{\infty} dx \left( \langle Q(x), X_{2s+1}\rangle - 2 \langle J, X_{2s+2}\rangle \right).
\end{split}\end{equation}
Since in our case $Q_2 =Q_4 =\cdots = 0$ we find that $X_{2s} \in \mathfrak{g}^{(0)}$  and $X_{2s+1} \in \mathfrak{g}^{(1)}$
and therefore $I_1 =I_3 =\cdots = 0$. In calculating the Lax pair for DNLS we in fact calculated the first four coefficients
$X_s=V_s$ for $s=1,\dots, 4$. Using this we get (see the appendix):
\begin{equation}\label{eq:Ik}\begin{aligned}
 I_1 &= 0, &\quad I_2 &=  \frac{1}{4} \int_{-\infty}^{\infty} dx \left( i \langle \q_x,\p \rangle - i \langle \q,\p_x \rangle +  \langle \q\p,\q\p \rangle \right), \\
 I_3 &= 0, &\quad I_4 &= \frac{1}{4} \int_{-\infty}^{\infty} dx \left( \langle \q_x,\p_x \rangle + \frac{i}{2} ( \langle \q_x,\p\q\p \rangle - \langle \q\p\q, \p_x \rangle)
+ \frac{1}{4}  \langle \q\p\q,\p\q\p \rangle \right).
\end{aligned}\end{equation}

\section{Conclusions}

We have studied quadratic bundle Lax pairs on $\mathbb{Z}_2$-graded Lie algebras and on {\bf A.III} symmetric spaces. This includes: the construction of Lax pairs
and the related NLEE of Kaup-Newell and GI type. We also constructed  the Jost solutions and the minimal set of scattering
data for local and nonlocal reductions. The later lead to equations having ${\cal CPT}$-symmetry.

We have also constructed the fundamental analytic solutions (FAS) and discussed briefly the spectral properties of the associated Lax operators.
It turns out that the spectral properties of the Lax operator depend crucially on the choice of representation of the underlying Lie algebra
or symmetric space while the minimal set of scattering data is provided by the same set of functions \cite{GG10}.

We have also formulated the Riemann-Hilbert problem for the Kaup-Newell (KN) and GI equations on {\bf A.III}-symmetric
spaces and derived explicit parametrization of the associated Lax operators. This can serve as a starting point in obtaining the Lax pair and the corresponding NLEEs \cite{107s,GYa*14}.

Finally, we have presented a modification of the dressing method and obtained 1-solitons for the
multi-component GI equation with local and nonlocal reductions. We have shown that for specific choices of
the polarization vector these solutions can develop singularities in finite time and that there are also cases
of soliton solutions with a regular behavior.

The results of this paper can be extended in several directions:
\begin{itemize}

\item To construct gauge covariant formulation of the multi-component KN and GI hierarchies on symmetric
spaces, including the generating (recursion) operator and it spectral decomposition, the description of the infinite set of
integrals of motion, the hierarchy of Hamiltonian structures.

\item To study the gauge equivalent systems to the multi-component KN and GI equations on symmetric spaces.

\item To study the associated Darboux transformations and their generalizations for DNLS equations over
Hermitian symmetric spaces and to obtain multi-soliton solutions via such generalizations. This includes also  rational solutions \cite{Ling,DokLeb,SuHe}.

\item To extend our results for the case of non-vanishing boundary conditions (a non-trivial
background). Such solutions were obtained  for the case of the scalar DNLS by using a slightly
different version of dressing method we have employed here \cite{Kawata1,Kawata2,Steudel,ZaMaLomi,MiZa*80}. In the scalar case, such solutions are of interest in nonlinear optics: they arise
in the theory of ultrashort femto-second nonlinear pulses in optical fibers, when
the spectral width of the pulses becomes comparable with the carrier frequency
and the effect of self-steepening of the pulse should be taken into account \cite{Chen}. The considerations required in this
case are more complicated and will be discussed it elsewhere.

  \item To study quadratic bundles associated
with other types of Hermitian symmetric spaces both for Kaup-Newell and for GI  equations \cite{20,76,GVYa*08,Rei,ReiSTSH}.

\end{itemize}

\section*{Acknowledgements}

The authors  thank Prof. V. V. Sokolov for the useful discussions.  RII acknowledges Seed
funding grant support from Dublin Institute of Technology for a project in association with ESHI Institute (Dublin).
Finally we thank the anonymous referees for the careful reading of the paper and  for
useful suggestions.

\appendix
\section{Derivation of the integrals of motion}

Let us introduce the notations:
\begin{equation}\label{eq:XX}\begin{split}
i \frac{\partial \xi}{ \partial x } \xi^{-1}(x,t,\lambda) \equiv i \frac{\partial \mathcal{Q}}{ \partial x } + \sum_{s=1}^{\infty} \frac{1}{(s+1)!}
\ad_{\mathcal{Q}}^s \frac{\partial \mathcal{Q}}{ \partial x } = \sum_{s=1}^{\infty} \lambda^{-s}\mathcal{X}_s, \\
\xi J \xi^{-1}(x,t,\lambda) \equiv J + \sum_{s=1}^{\infty} \frac{1}{s!} \ad_{\mathcal{Q}}^s J = J + \sum_{s=1}^{\infty} \lambda^{-s} X_s.
\end{split}\end{equation}
Then the fact that $\xi (x,t,\lambda)$ provide the FAS of the operator
\begin{equation}\label{eq:}\begin{split}
 i \frac{\partial \xi}{ \partial x } + (U_2(x,t) + \lambda U_1(x,t))\xi(x,t,\lambda) - \lambda^2 [J,\xi(x,t,\lambda)] =0,
\end{split}\end{equation}
leads to $U_1(x,t) \equiv Q(x,t)= [J, Q_1(x,t)]$, $U_2(x,t) = [Q_1(x,t), Q(x,t)]$ and to the recurrent relations
\begin{equation}\label{eq:RR}\begin{split}
i \mathcal{X}_s + X_{s+2} =0, \qquad s=1,2,\dots .
\end{split}\end{equation}
Obviously equations (\ref{eq:XX}) provide a recurrence to evaluate $\mathcal{X}_s$ and $X_s$ in terms of $Q_1(x,t)$.
Below we list the explicit formulae for the first few of them.
\begin{equation}\label{eq:A1}\begin{aligned}
 X_1 &= \ad_{Q_1}J = - U_1(x,t) =- \left(\begin{array}{cc} 0 & \q \\ \p & 0   \end{array}\right), &\;
  X_2 &= \frac{1}{2}\ad_{Q_1}^2J = - U_2(x,t)=-\frac{1}{2} \left(\begin{array}{cc}  \q\p & 0 \\ 0 & -\p \q  \end{array}\right). \\
\mathcal{X}_1 &=  \frac{\partial Q_1}{ \partial x } = \frac{1}{2}\left(\begin{array}{cc} 0 & \q_x \\ -\p_x &0  \end{array}\right), &\;
\mathcal{X}_2 &=  \frac{1}{2}\ad_{Q_1}\frac{\partial Q_1}{ \partial x } = \frac{1}{8}\left(\begin{array}{cc}  \q_x \p -\q\p_x & 0 \\
 0 &  \p_x \q -\p\q_x  \end{array}\right),
\end{aligned}\end{equation}
\begin{equation}\label{eq:A2}\begin{aligned}
X_3 &=  \ad_{Q_3}J +\frac{1}{6}\ad_{Q_1}^3J =-i \mathcal{X}_1, \qquad Q_3 = \frac{i}{4}\left(\begin{array}{cc}
0 & i \q_x + \frac{1}{3} \q\p\q \\ i\p_x - \frac{1}{3} \p\q\p & 0 \end{array}\right) , \\
 \mathcal{X}_3 &=  \frac{\partial Q_3}{ \partial x }+ \frac{1}{6}\ad_{Q_1}\frac{\partial Q_1}{ \partial x } =
 \frac{i}{4}\left(\begin{array}{cc}  0 & \q_{xx} \\ \p_{xx} & 0  \end{array}\right) + \frac{1}{16} \left(\begin{array}{cc}
0 & (\q\p\q)_x + \q\p_x \q \\ -(\p\q\p)_x - \p\q_x \p & 0  \end{array}\right).
\end{aligned}\end{equation}
\begin{equation}\label{eq:X4}\begin{split}
X_4 &= \frac{1}{2}\left( \ad_{Q_1} \ad_{Q_3} +\ad_{Q_3} \ad_{Q_1} \right) J + \frac{1}{24} \ad_{Q_1}^4 J \\
&= - \frac{1}{4}\left(\begin{array}{cc} i (\q_x\p -\q\p_x) + \frac{1}{2} \q\p\q\p & 0 \\ 0 & i(\p_x\q - \p\q_x) - \frac{1}{2} \p\q\p\q \end{array}\right) , \\
X_5 &=   \ad_{Q_5}J+ \frac{1}{6}  \left( \ad_{Q_1}^2 \ad_{Q_3} +  \ad_{Q_1}  \ad_{Q_3} \ad_{Q_1} + \ad_{Q_3} \ad_{Q_1}^2 \right) J + \frac{1}{120} \ad_{Q_1}^5 J \\
&= \frac{1}{4} \left(\begin{array}{cc} 0 & \q_{xx} - \frac{i}{4}(\q\p\q)_x - \frac{i}{4} \q\p_x\q \\ \p_{xx} + \frac{i}{4} (\p\q\p)_x +\frac{i}{4} \p\q_x\p & 0
 \end{array}\right), \\
 Q_5 &= \frac{1}{8} \left(\begin{array}{cc} 0 & -\q_{xx} + \frac{7i}{12} (\q\p\q)_x - \frac{5i}{12} \q\p_x\q + \frac{3}{10} \q\p\q\p\q \\
\p_{xx} + \frac{7i}{12} (\p\q\p)_x - \frac{5i}{12} \p\q_x\p + \frac{3}{10} \p\q\p\q\p & 0   \end{array}\right).
\end{split}\end{equation}
The above expressions readily lead to the results for the conserved quantities (\ref{eq:Ik}) along with the fact that
\begin{equation}\label{eq:JX}\begin{split}
\langle J, \xi J \xi^{-1}(x,t,\lambda)\rangle = \langle J, J \rangle + \sum_{s=1}^{\infty} \frac{(-1)^s}{(2s)!}
\left\langle  \ad_{\mathcal{Q}}^s J, \ad_{\mathcal{Q}}^s J \right\rangle .
\end{split}\end{equation}


\begin{thebibliography}{99}


\bibitem{Konotop} F.K. Abdullaev, Y.V. Kartashov, V.V. Konotop, D.A. Zezyulin, {\it Solitons in PT-symmetric
nonlinear lattices}, Phys. Rev. A {\bf 83} (2011), 041805.

\bibitem{AKNS}
Ablowitz M.J., Kaup D.J., Newell A.C., Segur H.,
{\it The inverse scattering transform -- Fourier analysis for nonlinear problems},
 Studies in Appl. Math.  {\bf 53} (1974), 249--315.

\bibitem{AblMus} M. J. Ablowitz and Z. H. Musslimani, {\it Integrable Nonlocal Nonlinear Schr\"odinger
Equation}, Phys. Rev. Lett., {\bf 110} (2013) 064105.

\bibitem{AblMus1} M. J. Ablowitz and Z. H. Musslimani, {\it Integrable discrete ${\cal PT}$ symmetric model}, Phys. Rev. E {\bf 90} (2014) 032912.

\bibitem{AblMus2} M. J.  Ablowitz and Z. H.  Musslimani, {\it Inverse scattering transform for the integrable
nonlocal nonlinear Schr\"odinger equation}, Nonlinearity {\bf 29} (2016) 915--946.

\bibitem{AthFor}
Athorne C., Fordy A., {\it Generalised KdV and MKdV equations associated with symmetric spaces}, J.~Phys.~A: Math. Gen. {\bf 20} (1987), 1377--1386.

\bibitem{Barash} I. V. Barashenkov, {\it Hamiltonian formulation of the standard ${\cal PT}$-symmetric
nonlinear Schr\"odinger dimer}, Physical Review A 90 (2014) 045802.

\bibitem{Barash1} I V Barashenkov, D E Pelinovsky and P Dubard, {\it Dimer with gain and loss: Integrability
 and ${\cal PT}$-symmetry restoration}, J. Phys. A: Math. Theor. {\bf 48} (2015) 325201.

\bibitem{Bender1} C. M. Bender and S. Boettcher, {\it Real Spectra in Non-hermitian Hamiltonians
Having ${\cal PT}$ Symmetry}, Phys. Rev. Lett 80 (1998) 5243--5246;\\
    C. M. Bender, S. Boettcher and P. N. Meisinger, {\it ${\cal PT}$-Symmetric Quantum Mechanics},  J. Math. Phys. {\bf 40} (1999) 2201--2229.

\bibitem{Bender2}  C. M. Bender, {\it Making Sense of Non-hermitian Hamiltonians},  Rep. Progr. Phys. {\bf 70} (2007) 947--1018 (E-print: {\tt hep-th/0703096}).

\bibitem{Chen} X. Chen and W. Lam, {\it Inverse scattering transform for the derivative nonlinear Schr\"odinger
equation with nonvanishing boundary conditions}, Phys. Rev. E {\bf 69} (2004) 066604.

\bibitem{Chered} I. V. Cherednik, {\it Factorizing particles on a half-line and root systems}, Theor. Math. Phys. {\bf 61} (1984), 977--983.

\bibitem{DokLeb}  {E. V. Doktorov, and  S. B. Leble}, {\it Dressing method in Mathematical Physics}, Mathematical Physics Studies {\bf 28},
Springer, Berlin, Heidelberg, New York (2007).

\bibitem{DriSok} V.V. Drinfel'd and  V. V. Sokolov,  {\it Lie algebras and equations of Korteweg-de Vries type}, Journal of Soviet Mathematics, {\bf 30} (1985) 1975--2036.

\bibitem{FaTa} L.~D.~Faddeev, L.~A.~Takhtadjan, {\it Hamiltonian
Method in the Theory of Solitons}, Springer Verlag, Berlin (1987).

\bibitem{Fan} E. Fan,  {\it Darboux transformation and soliton-like solutions for the Gerdjikov-Ivanov equation}, J. Phys. A: Math. Gen. {\bf 33}
(2000) 6925--6933.

\bibitem{For}
Fordy A.P., {\it Derivative nonlinear Schr\"odinger equations and Hermitian symmetric spaces}, J. Phys.~A: Math. Gen. {\bf 17}  (1984), 1235--1245.

\bibitem{ForKu*83}
Fordy A.P., Kulish P.P., {\it Nonlinear Schr\"{o}dinger equations and simple Lie algebras},  Comm. Math. Phys. {\bf 89} (1983), 427--443.

\bibitem{Fring1} A Fring, {\it ${\cal PT}$-symmetric deformations of the Korteweg-de Vries equation}, J. Phys. A: Math. Theor. {\bf 40} (2007), 4215 (E-print: {\tt math-ph/0701036}).

\bibitem{Fring2} A Fring, {\it ${\cal PT}$-symmetric deformations of integrable models}, Phil. Trans. Royal Soc. A {\bf 371} (2013), 20120046.

\bibitem{GeDik} I. M. Gelfand  and  L. A. Dickey, {\it The resolvent and Hamiltonian systems}, Funct. Anal. Appl. {\bf 11} (1977) 93--105.


\bibitem{IP2}  {V. S. Gerdjikov},  {\it Generalised Fourier transforms for  the  soliton
equations. Gauge covariant formulation}, Inverse Problems {\bf 2} (1986), 51--74.

\bibitem{67} V. S. Gerdjikov, {\em Algebraic and Analytic Aspects of $N $-wave Type Equations}, Contemporary Mathematics {\bf 301} (2002) 35--68
(E-print: {\tt nlin.SI/0206014}).

\bibitem{107s} V. S. Gerdjikov, {\it Riemann-Hilbert problems with canonical normalization
and families of commuting operators},   Pliska Stud.\ Math.\ Bulgar.\  {\bf 21} (2012), 201--216 (E-print: {\tt arXiv:1204.2928}).

\bibitem{GG10} V. S. Gerdjikov and G. G. Grahovski, {\it  Multi-Component NLS Models on Symmetric Spaces:
Spectral Properties versus Representations Theory}, Symmetry, Integrability and Geometry:
Methods and Applications (SIGMA) {\bf 6} (2010), paper 044 (29 pages) (E-print: {\tt nlin.SI/1006.0301}).

\bibitem{GGK*XX}  V. S. Gerdjikov, G. G. Grahovski,  and R. I. Ivanov,
{\it On the $N$-wave  equations with ${\cal PT}$-symmetry}, Theor. Math. Phys. {\bf 188} (2016), 1305--1321 (E-print: {\tt arXiv:1601.01929}).

\bibitem{vgrn} V. S. Gerdjikov, G. G. Grahovski, R. I. Ivanov, N. A. Kostov,
{\it $N $-wave interactions related to simple Lie algebras.
${\Bbb Z}_2$- reductions and soliton solutions}, Inverse Problems {\bf 17} (2001) 999--1015
(E-print: {\tt nlin.SI/0009034}).

\bibitem{GGK*01}  {V. S. Gerdjikov, G. G. Grahovski,  and N. A. Kostov},
{\it Reductions of $N $-wave interactions related to low--rank simple Lie algebras. I: ${\Bbb Z}_2$- reductions},
J. Phys. A: Math \& Gen. {\bf 34}, 9425--9461 (2001) (E-print: {\tt nlin.SI/0006001}).

\bibitem{GGK*01b}  {V. S. Gerdjikov, G. G. Grahovski,  and N. A. Kostov},
{\it On $N$-wave type systems and their gauge equivalent}, European  Physical Journal B {\bf 29} (2002), 243--248 (E-print: {\tt nlin.SI/0111027}).

\bibitem{GGK05a} V.~S.~Gerdjikov, G.~G.~Grahovski  and N.~A.~Kostov, {\it On the
multi-component NLS type equations on symmetric spaces and their
reductions}, Theor. Math. Phys. {\bf 144} (2005), No.2, 1147--1156.

\bibitem{20}  {V. S. Gerdjikov,  and M. I. Ivanov},
{\it The quadratic bundle of  general  form  and  the
nonlinear evolution equations.  I.  Hierarchies of Hamiltonian structures}, Bulgarian J. Phys. {\bf 10} (1983) No.1, 13--26  [In Russian].

\bibitem{21}  {V. S. Gerdjikov,  and M. I. Ivanov},
{\it The quadratic bundle of  general  form  and  the
nonlinear evolution equations.  II.  Hierarchies of Hamiltonian structures}, Bulgarian J. Phys. {\bf 10} (1983) No.2, 130--143  [In Russian].

\bibitem{76} V. S. Gerdjikov, M. I. Ivanov, P. P. Kulish, {\it Quadratic bundle and nonlinear equations}, Theor. Math. Phys. {\bf 44} (1980), 784--795.

\bibitem{GKKV*08}  {V. S. Gerdjikov, D. J. Kaup, N. A. Kostov,  and T. I. Valchev.}
On classification of soliton solutions of multicomponent nonlinear evolution equations.
\textit{J. Phys. A: Math. Theor.} {\bf 41} (2008) 315213 (36pp).

\bibitem{GK*81} V.~S.~Gerdjikov,  and P.~P.~Kulish.
{\it    The generating operator for the $n \times n$ linear system.  }
    Physica  {\bf 3D,} n.~3, 549--564, (1981).

\bibitem{GVYa*08}  V. S. Gerdjikov, G. Vilasi,  and A. B. Yanovski, {\it Integrable Hamiltonian Hierarchies. Spectral and Geometric Methods}, Lecture Notes in Physics {\bf 748},
Springer Verlag, Berlin, Heidelberg, New York (2008).

\bibitem{GYa*14}  V. S. Gerdjikov   and A. B. Yanovski, {\it Riemann-Hilbert problems, families of commuting
operators and soliton equations}, J. Phys: Conf. Series {\bf 482} (2014) 012017.

\bibitem{Grah} G. G. Grahovski, {\it The Generalised Zakharov-Shabat System and the Gauge Group Action}, J. Math. Phys.  {\bf 53} (2012)  073512 (E-print: {\tt nlin.SI/1109.5108}).

\bibitem{Helg}
Helgason S., {\it Differential geometry, Lie groups and Symmetric
Spaces}, Graduate Studies in Mathematics {\bf 34}, AMS, Providence, Rhode Island (2001).

\bibitem{RI}  {R. Ivanov.} On the dressing method for the
generalised Zakharov--Shabat system, {\it Nuclear Phys.~B} {\bf 694} (2004); 509--524, (E-print: {\tt math-ph/0402031}).

\bibitem{KN} D. J. Kaup and A. C. Newell, {\it An exact solution for a derivative nonlinear Schr\"odinger equation}, J. Math. Phys. {\bf 19} (1978), 798--801.

\bibitem{Kawata1} T. Kawata and H. Inoue, {\it Exact solutions of the derivative nonlinear Schr\"odinger equation under the nonvanishing conditions}, J. Phys. Soc. Japan {\bf 44} (1978) 1968--1976.

\bibitem{Kawata2} T. Kawata, J. Sakai and N. Kobayashi, {\it Inverse method for the mixed nonlinear Schr\"odinger equation and soliton solutions}, J. Phys. Soc. Japan {\bf 48} (1980) 1371--1379.

\bibitem{Khabib} I. T. Khabibullin, {\it Boundary conditions for nonlinear equations compatible with integrability}, Theor. Math. Phys. {\bf 96} (1993), 845--853.

\bibitem{Kundu1} A. Kundu, {\it Exact solutions to higher-order nonlinear equations through gauge transformation}, Physica D {\bf 25} (1987) 399--406.

\bibitem{Kundu2} A. Kundu, {\it Integrability of classical and semiclassical derivative non-linear Schr\"odinger equation with
non-ultralocal canonical structure}, J. Phys. A: Math. Gen. {\bf 21} (1988) 945--953.

\bibitem{KuRei}  {P. P. Kulish,  and A. G. Reiman}, {\it Hamiltonian structure of polynomial bundles}, Sci. Notes of LOMI seminars {\bf 123} (1983) 67--76.

\bibitem{KuzMikh}     E. A. Kuznetsov, A. V. Mikhailov, {\it On the complete integrability of the two-dimensional classical Thirring model}, Theor. Math. Phys. {\bf 30} (1977) 193--200.

\bibitem{Lee}
Y. C. Lee, H. H. Chen and C. S. Liu, {\it Integrability of Nonlinear Hamiltonian Systems by Inverse Scattering Method}, Phys. Scr. {\bf 20} (1979) 490--492.

\bibitem{Ling} L. Ling and Q. P, Liu, {\it Darboux transformation for a two-component
derivative nonlinear Schr\"odinger equation}, J. Phys. A: Math. Theor. {\bf 43} (2010) 434023 (11pp).

\bibitem{2}  A. V. Mikhailov,  {\it The reduction problem and the inverse
scattering problem},   Physica D {\bf 3} (1981), 73--117.

\bibitem{Mio} K. Mio, T. Ogino, K. Minami  and S. Takeda, {\it Modified Nonlinear Schr\"odinger Equation for Alfv\'en
Waves Propagating along the Magnetic Field in Cold Plasmas}, J. Phys. Soc. Japan  {\bf 41} (1976), 265--271.

\bibitem{Mjo} E. Mj{\o}lhus, {\it On the modulational  instability  of  hydromagnetic waves parallel to  the magnetic field}, J.  Plasma  Physics
  {\bf 16} (1976),   321--334.

\bibitem{Ali1}  A. Mostafazadeh,    {\it Pseudo-hermiticity    versus     ${\cal PT}$-Symmetry I, II, III}, J. Math. Phys.
{\bf 43} (2002) 205--214 (E-print: {\tt math-ph/0107001}); 2814--2816 (E-print: {\tt math-ph/0110016}); 3944--3951 (E-print: {\tt math-ph/0203005}).

\bibitem{Ali2} A. Mostafazadeh, {\it Pseudo-hermiticity and Generalized ${\cal PT}$- and ${\cal CPT}$-Symmetries},
J. Math. Phys. {\bf 44} (2003) 974--989 (E-print: {\tt math-ph/0209018});\\
    A. Mostafazadeh, {\it Exact ${\cal PT}$-Symmetry Is Equivalent to Hermiticity}, J. Phys. A: Math. Gen.
    {\bf 36} (2003) 7081--7091 (E-print: {\tt quant-ph/0304080}).


\bibitem{ZMNP}   { S. P. Novikov, S. V. Manakov, L. P. Pitaevsky,  and V. E. Zakharov.} {\it Theory of solitons: the inverse scattering method.}
 Plenum, New York, (1984).

\bibitem{Peskin} M. E. Peskin, D. V. Schroeder, {\it An introduction to Quantum Field Theory}, Perseus Books, Reading, Massachusetts (1995).

\bibitem{Rei}  {A. G. Reiman.}
A unified Hamiltonian system on polynomial bundles, and
the structure of stationary problems.
\textit{Sci. Notes of LOMI seminars} vol. {\bf 131}, pp. 118--127, (1983).

\bibitem{ReiSTSH}  {A.G. Reiman,  and M. A. Semenov-Tyan-Shanskii.} Current algebras and nonlinear partial
differential equations, \textit{Dokl. Akad. Nauk SSSR,} {\bf 251,} 1310--1312 (1980).

\bibitem{Ruder1} M. S. Ruderman, {\it DNLS equation for large-amplitude solitons propagating in an arbitrary direction in a high-$\beta$ Hall plasma}, J. Plasma Physics {\bf 67} (2002), 271--276.

    \bibitem{Ruder2} M. S. Ruderman, {\it Freak waves in laboratory and space plasmas}, Eur. Phys. J. (Spec. Top.) {\bf 185} (2010), 57--66.

\bibitem{Nature}  C. E.  R\"uter,  K. G.  Makris,  R.  El-Ganainy,  D. N. Christodoulides, M. Segev and D. Kip,
{\it Observation of parity-time symmetry in optics}, Nature Physics {\bf 6} (2010) 192 -- 195.

\bibitem{Sh}  Shabat A. B.
{\it The inverse scattering problem for a system of differential
equations.} Functional Annal. \& Appl. {\bf 9}, n.3, 75--78 (1975);\\
 Shabat A. B. {\it The inverse scattering problem.}  Diff. Equations
{\bf 15}, 1824--1834 (1979).

\bibitem{Sklyan} E. K. Sklyanin, {\it Boundary conditions for integrable equations}, Funct. Anal. Appl. {\bf 21} (1987), 164--166.


\bibitem{Steudel} H. Steudel, {\it The hierarchy of multi-soliton solutions of the derivative nonlinear Schr\"odinger equation}, J. P
hys. A: Math. Gen. {\bf 36} (2003) 1931--1946.

\bibitem{SuHe} Sh. Su, J. He and L. Wang, {\it The Darboux transformation of the derivative nonlinear Schr\"odinger equation}, J. Phys. A: Math. Theor.
{\bf 44} (2011) 305203.

\bibitem{Tsuchida} T. Tsuchida and M. Wadati, {\it Complete integrability of derivative nonlinear Schr\"odinger-type equations}, Inv. Problems {\bf 15}
(1999) 1363--1373.

\bibitem{TV0}
T. I.  Valchev, {\it On the Quadratic Bundles Related to Hermitian Symmetric Spaces},  J. Geom. Symmetry Phys. {\bf 29} (2013), 83--110.

\bibitem{TV}
T. I.  Valchev, {\it On Mikhailov's reduction group}, Phys. Lett A
{\bf 379} (2015)  1877--1880.

\bibitem{TV2}
T. I.  Valchev, {\it Dressing method and quadratic bundles related to symmetric spaces. Vanishing boundary conditions}, J. Math. Phys.
{\bf 57} (2016)  021508 (14 pp.).

\bibitem{Yilmaz} H. Yilmaz, {\it Exact solutions of the Gerdjikov-Ivanov equation using Darboux transformations}, J. Nonlim. Math. Phys. {\bf 22} (2015) 32-46.


\bibitem{ZaMaLomi}  {V. E. Zakharov,  and S. V. Manakov.} Multidimensional nonlinear integrable systems and methods for constructing
their solutions. \textit{Sci. Notes of LOMI seminars}, {\bf vol. 133}, pp. 77-91, (1984).

\bibitem{MiZa*80}     V. E. Zakharov and A. V. Mikhailov, {\it On the integrability of classical spinor models in
two-dimensional space--time},  Commun. Math. Phys. {\bf 74,} 21--40 (1980).

\bibitem{ZaSh*74a}  {V. E. Zakharov,   and A. B. Shabat.} "A scheme for integrating nonlinear evolution equations
of mathematical physics by the inverse scattering method. I," Funkts. Anal. Prilozhen.,
{\bf 8}, \textsl{No. 3,} 43--53 (1974).


\bibitem{ZaSh*79}   {V. E. Zakharov,  and A. B. Shabat.} Integration of the nonlinear equations of mathematical
physics by the inverse scattering method II, \textit{Funkts. Anal. Prilozhen.}, {\bf 13,} \textsl{No. 3,} 13--22 (1979).

\bibitem{UFN}  A. A. Zyablovsky, A. P. Vinogradov, A. A. Pukhov, A. V. Dorofeenko and A. A Lisyansky,
{\it ${\cal PT}$-symmetry in optics},  Phys.-Uspekhi {\bf 57} (2014), no. 11, 1063--1082.




\end{thebibliography}
\end{document}